\documentclass[twocolumn,english]{IEEEtran}
\usepackage[T1]{fontenc}
\usepackage[utf8]{inputenc}
\usepackage{geometry}
\usepackage{color}
\usepackage{amsmath}
\usepackage{amssymb}
\usepackage{graphicx}
\usepackage{esint}
\usepackage{caption}
\usepackage{subcaption}

\makeatletter

\providecommand{\tabularnewline}{\\}

\renewcommand{\fnum@figure}{Fig.~\thefigure}
\usepackage{cite}

\usepackage{babel}

\usepackage{booktabs}
\usepackage{arydshln}

\usepackage{babel}
\usepackage{xcolor}
\providecolor{lyxadded}{rgb}{0,0,1}
\providecolor{lyxdeleted}{rgb}{1,0,0}

\usepackage{pdfcolmk}
\usepackage{changes}
\usepackage{xcolor}

\definecolor{rgb}{RGB}{0, 180, 0}

\usepackage{babel}

\makeatother

\usepackage{babel}
\begin{document}

\title{Solution of Cavity Resonance and Waveguide Scattering Problems Using
the Eigenmode Projection Technique}

\author{Mamdouh H. Nasr, \textit{Student Member, IEEE, }Mohamed A. K. Othman,
\textit{Student Member, IEEE, }Islam A. Eshrah, \textit{Senior Member,
IEEE,} and Tamer M. Abuelfadl\textit{, Member}, \textit{IEEE}\thanks{M. H. Nasr is with the Department of Electrical Engineering, Stanford
University, USA (mamdouh@stanford.edu).

M. A. K. Othman is with the Electrical Engineering and Computer Science
Department, University of California, Irvine, USA (mothman@uci.edu).

I. A. Eshrah and T. M Abuelfadl are with the Electronics and Electrical
Communication Engineering Department, Faculty of Engineering, Cairo
University, Giza, Egypt (ieshrah@eng.cu.edu.eg and tamer@eng.cu.edu.eg).}}
\maketitle
\begin{abstract}
An eigenmode projection technique (EPT) is developed and employed
to solve problems of electromagnetic resonance in closed cavities
and scattering from discontinuities in guided-wave structures. The
EPT invokes the eigenmodes of a canonical predefined cavity in the
solution procedure and uses the expansion of these eigenmodes to solve
Maxwell's equations, in conjunction with a convenient choice of port
boundary conditions. For closed cavities, resonance frequencies of
arbitrary-shaped cavities are accurately determined with a robust
and efficient separation method of spurious modes. For waveguide scattering
problems, the EPT is combined with the generalized scattering matrix
approach to solve problems involving waveguide discontinuities with
arbitrary dielectric profiles. Convergence studies show stable solutions
for a relatively small number of expansion modes, and the proposed
method shows great robustness over conventional solvers in analyzing
electromagnetic problems with inhomogeneous materials.\end{abstract}


\section{Introduction}

Waveguides and cavity resonators are among the oldest structures employed
in the microwave regime due to the low losses of these structures
at microwave frequencies. Bulky as they are, conducting waveguides
and cavities remain the most robust and reliable structures in the
realization of a broad spectrum of components and subsystems ranging
from filters, isolators, circulators, duplexers, to particle accelerators,
slotted waveguide arrays... etc. Modal solution to waveguide systems
and scatterers was widely explored in the literature \cite{ModalanalysisWg,modalanalysswgdiscont,modelexpansionclosed,planewaveepnasions}.
Hybrid methods were also investigated, wherein modal analysis is invoked
to evaluate the cutoff frequencies of dielectric-loaded waveguides
\cite{dielectric2Dwaveguide,wgdiscondielectric} or mode-matching-like
approaches are invoked to evaluate the expansion coefficients, along
with the use of a conventional numerical technique such as the finite-difference
time-domain (FDTD) method \cite{FTDTBCs}, the finite element method
(FEM) \cite{FEMBCs}, and the method of moments (MoM) \cite{hybridmommodematching}.
In this domain, several ideas were proposed to present a full wave
solution to scattering problems. In \cite{BIM_cav}
and \cite{BIM_wg}, a Boundary Integral Method (BIM) was proposed
to solve problems of cavity resonance and waveguide scattering based
on the Green's function of a spherical resonator approximated by a
rational function of frequency to eliminate frequency dependence of
the matrix solution. This was, however, limited to lossless, isotropic
and homogeneous media, in addition to the difficulties imposed by
the use of the Green's function, the need of using suitable basis
functions (such as the RWG bases) upon discretizing the surface, and
the involved calculations necessary to determine the generated matrices.

In the fifties of the last century, electromagnetic field expansion
inside cavities was introduced by Slater \cite{slatterbokk}. In his
work, the solenoidal and irrotational cavity eigenmodes, which form
a complete set, were used to represent vector fields inside microwave
cavities targeting the equivalent network modeling of microwave cavities
with output waveguide ports. Kurokawa \cite{kurokawacavity} generalized
the analysis after including irrotational modes in the magnetic field
representation. Several contributions were made to fully understand
the expansion and its properties \cite{schkonofimpedance,cavityteichman,schelknoffexpansion}.
Applications of such work were numerous in the field of microwave
filters which involve cavities as high quality factor resonators,
and especially in high power applications \cite{filterhighpower}
and particle accelerators \cite{slater1948design}.

However, this strong theoretical framework has not been hitherto utilized
to solve the sophisticated problems of scattering from arbitrary objects
within the waveguide environment. Recently, the Eigenmode Projection
Technique (EPT) was proposed to solve different scattering problems
in guided (closed) and unguided (open) structures \cite{myAPS,Mam_APS,Mam_EuMC,othman_IMS2013,Mam_journal}
due to its versatility, fast convergence, and straightforwardness
implementation. It relies on the expansion of fields in terms of canonical
(analytically known) complete set of modes to represent fields in
arbitrarily shaped geometries, without the need of prior discretization
of the solution domain \cite{Mam_journal}. In this paper, the proposed
EPT benefits from the completeness of the eigenmodes of canonical
cavities and the corresponding expansion to find a modal solution
to Maxwell's equations using mode projection, i.e., retrieve the values
for the unknown fields expansion coefficients to construct the full
wave solution for resonance and scattering problems in guided structures.
The Generalized Scattering Matrix (GSM) approach facilitates the evaluation
of the equivalent terminal voltages and currents, and the corresponding
S-parameters \cite{impednormhjaskal} in the case of existing excitation
ports. The strength of the presented technique lies
in the fact that it (a) is fully capable of treating non-canonical
problems using the same procedure of treating canonical cavities,
(b) lends itself to integration with other techniques, such as the
GSM, in a natural way, (c) is applied not only to closed problems
(the scope of this work) but also to open/scattering problems, (d) lends itself inherently to problems involving bodies of revolution (BoRs), and
(e) is easily extended to periodic structures \cite{IMS2016_Tarek},
electrostatics \cite{IMS2016_Static}, and even full-blown 3D geometries.
For the sake of conciseness, the focus of this paper is on problems
of scattering and resonance in guided-wave structures. In Section
II, the formulation and the solution procedure for arbitrary electromagnetic
problems using eigenmode projections is detailed. Section III covers
the eigenmode solution of arbitrary shaped cavity, while Section IV
explores the scattering problems inside waveguides.

\section{Formulation of the Eigenmode Projection Technique}
\begin{figure}[t]
\centering
\includegraphics[width=9cm]{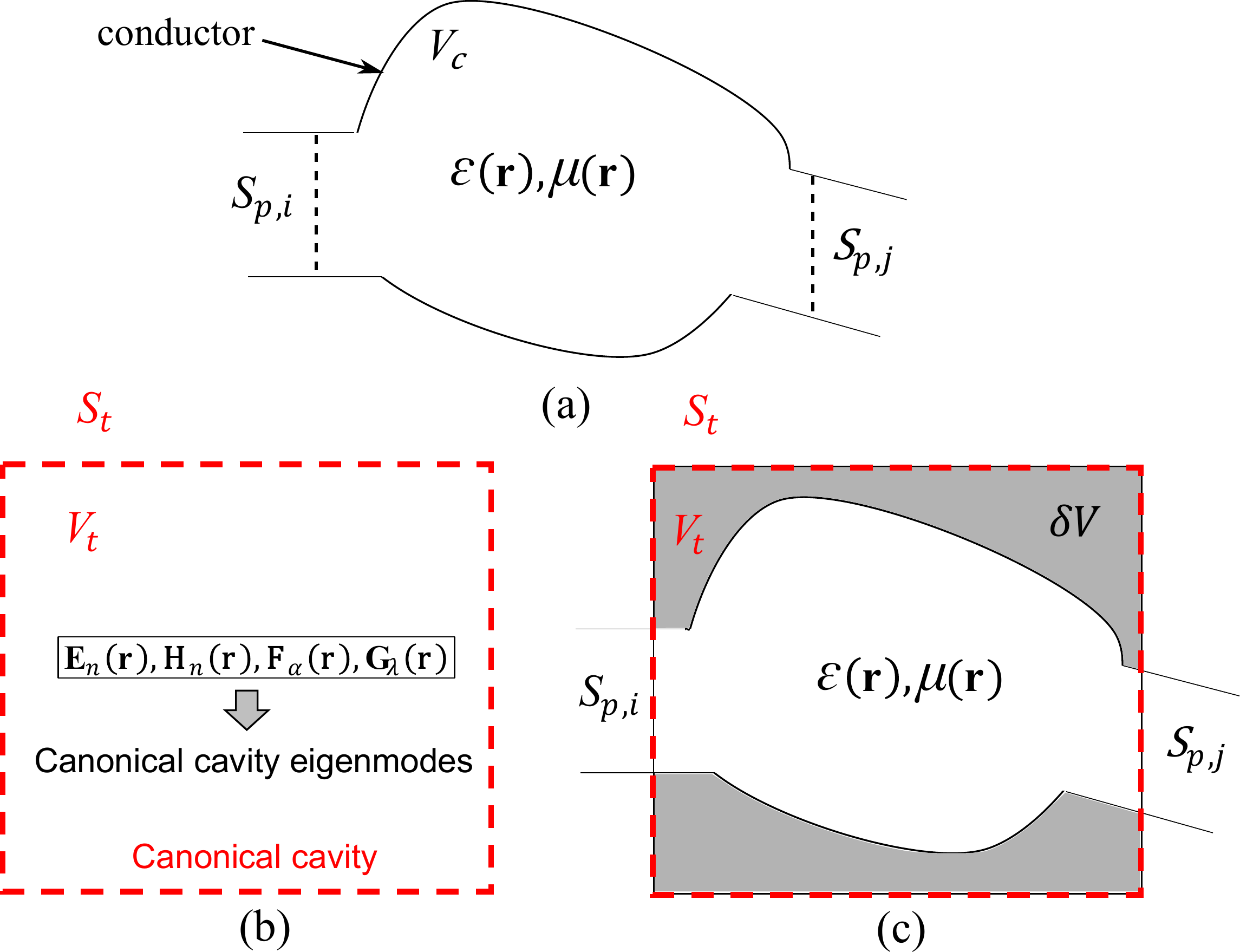}
\medskip{}
\includegraphics[scale=0.65]{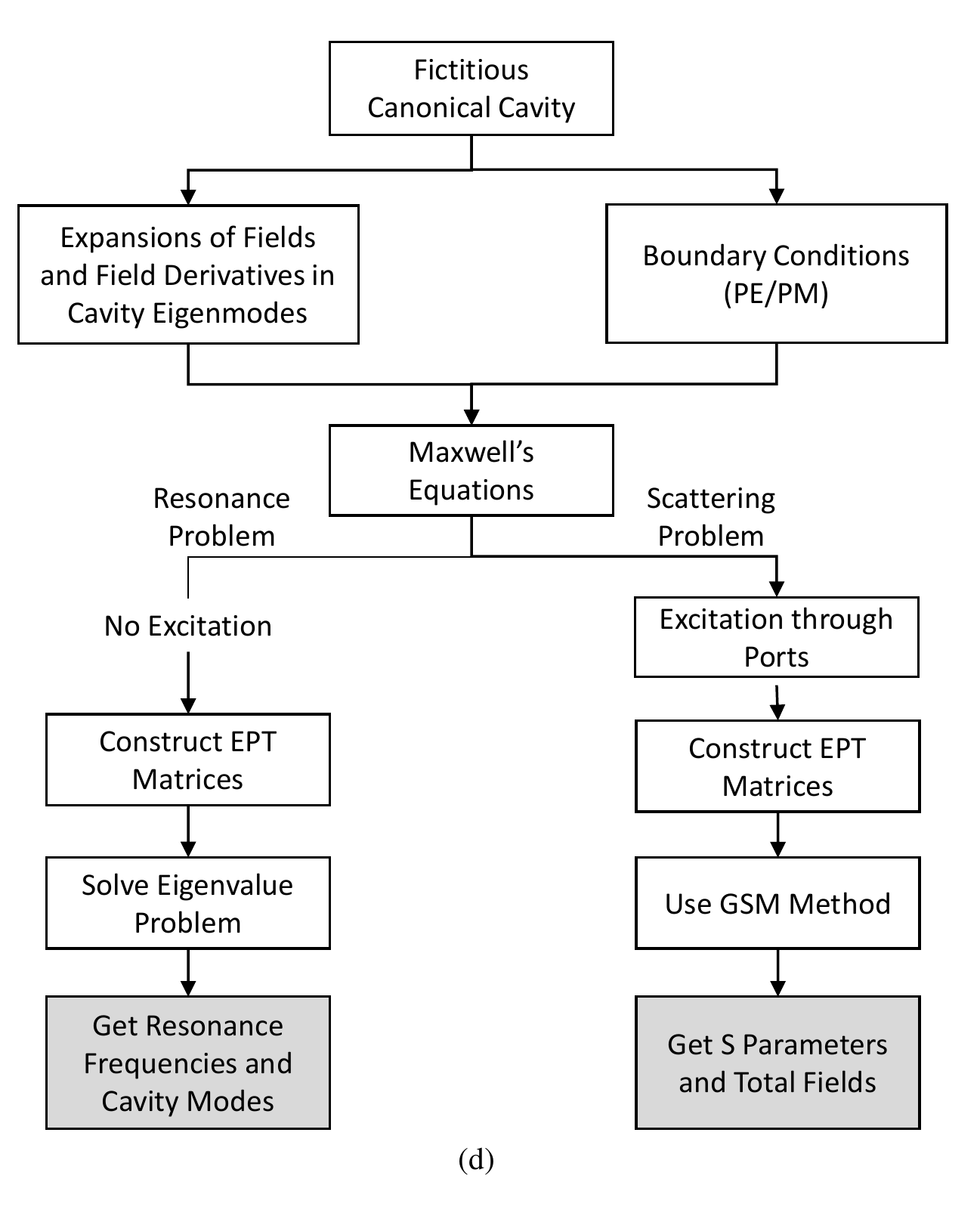}
\caption{\label{Config}(a) An arbitrary-shaped conducting cavity with arbitrary
material loading and multiple waveguide ports. (b) Applying the EPT
by utilizing canonical conducting cavity with known eigenmodes that
are used in expanding arbitrary fields inside. (c) The wave solution
of the problem in (a) is represented in terms of the canonical modes
in (b). (d) Flow chart of the solution procedure of the problem.}
\end{figure}

Consider the arbitrarily shaped conducting cavity shown in Fig \ref{Config}(a),
which is generally filled with arbitrary material $\varepsilon_{r}(\mathbf{r})$
and $\mu_{r}(\mathbf{r})$, and could be excited through a number
of waveguides with ports denoted by $S_{p}$. In order to solve such
a problem (i.e., finding a unique full-wave solution everywhere),
a canonical closed cavity that encapsulates the arbitrary-shaped cavity,
with enclosing surfaces either perfect electric (PE) or perfect magnetic
(PM) is chosen (as illustrated in Fig. \ref{Config}(b) with dashed
red color) within which the fields can be expanded in terms of known
solenoidal and irrotational eigenmodes. The field expansions are used
in Maxwell’s equations to solve inside the canonical cavity in presence
of those materials. Subsequently, mode projections are performed.
If external sources are present, parts of the fictitious cavity surface
are regarded as waveguide ports (Fig. \ref{Config}(c)), and appropriate
boundary conditions are chosen and enforced to address the coupling
to the external fields. Then, an equivalent microwave network is developed
where the GSM approach is employed. Finally, the field coefficients
are retrieved by solving a system of linear equations, conveniently
cast in matrix form and subsequently solved using a suitable technique.
A flow-chart representing the
solution procedure is given in Fig. \ref{Config}(d).

\subsection{Canonical Cavity Eigenmode Expansion}
Consider a canonical domain $V_{t}$ bounded by a connected surface
$S_{t}$ as depicted in Fig. \ref{Config}(b). The electric and magnetic
fields may be cast in the form of an expansion of a known complete
set of orthogonal solenoidal (divergence-free) and irrotational (curl-free)
eigenmodes \cite{slatterbokk,vbladelbook} as
\begin{gather}
\mathbf{E}(\mathbf{r})=\sum_{n}a_{n}\mathbf{E}_{n}(\mathbf{r})+\sum_{\alpha}f_{\alpha}\mathbf{F}_{\alpha}(\mathbf{r}),\label{eq:e_cav}\\
\mathbf{H}(\mathbf{r})=\sum_{n}b_{n}\mathbf{H}_{n}(\mathbf{r})+\sum_{\lambda}g_{\lambda}\mathbf{G}_{\lambda}(\mathbf{r}).\label{eq:h_cav}
\end{gather}
The solenoidal eigenmodes ($\mathbf{E}_{n}$, $\mathbf{H}_{n}$) are
coupled through the curl equations
\begin{gather}
\nabla\times\mathbf{E}_{n}=k_{n}\mathbf{H}_{n},\,\nabla\times\mathbf{H}_{n}=k_{n}\mathbf{E}_{n},\label{eq:curl_solenoidal}
\end{gather}
and satisfy the homogeneous Helmholtz equation, viz.
\begin{gather}
\left(\nabla^{2}+k_{n}^{2}\right)\mathbf{E}_{n}=0,\,\left(\nabla^{2}+k_{n}^{2}\right)\mathbf{H}_{n}=0.\label{eq:Helmoltz_solenoida}
\end{gather}
On the other hand, the irrotational eigenmodes ($\mathbf{F_{\alpha}},\,\mathbf{G_{\lambda}}$),
are represented by the gradient of the scalar potentials ($\Phi_{\alpha}$,
$\Psi_{\lambda}$) through
\begin{gather}
l_{\alpha}\mathbf{F_{\alpha}}=\nabla\Phi_{\alpha},\,w_{\lambda}\mathbf{G_{\lambda}}=\nabla\Psi_{\lambda},\label{eq:grad_irr}
\end{gather}
with those scalar potentials satisfying, in turn, Helmholtz equation
\begin{gather}
\left(\nabla^{2}+l_{\alpha}^{2}\right)\Phi_{\alpha}=0,\,\left(\nabla^{2}+w_{\lambda}^{2}\right)\Psi_{\lambda}=0,\label{eq:Helmoltz_irr}
\end{gather}
where $k_{n},\,l_{\alpha},\,w_{\lambda}$ are the wavenumbers (eigenvalues)
for the solenoidal and irrotational electric and magnetic fields,
respectively. Needless to say, the infinite number of modes is truncated
to $N$ solenoidal modes and $M,\,K$ irrotational electric and magnetic
ones, respectively. An example for the solution of the above equations is detailed
in the Appendix for the encountered sample problems.

In order to use those expansions in Maxwell's equations, expansions
for $\nabla\times\mathbf{E}$, $\nabla\times\mathbf{H}$, $\nabla\cdot\mathbf{D}$
and $\nabla\cdot\mathbf{B}$ are required, but cannot be obtained
by directly applying the curl and divergence operators to (\ref{eq:e_cav})
and (\ref{eq:h_cav}). Careful algebraic manipulations (with details
in \cite{collinbookfoundation,kurokawacavity,vbladelbook}) yield
\begin{multline}
\nabla\times\mathbf{E}=\sum_{n}\left(k_{n}a_{n}+\oint_{S_{t}}\mathbf{E}\times\mathbf{H}_{n}\cdot d\mathbf{s}\right)\mathbf{H}_{n}\\
+\sum_{\lambda}\left(\oint_{S_{t}}\mathbf{E}\times\mathbf{G}_{\lambda}\cdot d\mathbf{s}\right)\mathbf{G}_{\lambda}\label{eq:curl_e}
\end{multline}
\begin{multline}
\nabla\times\mathbf{H}=\sum_{n}\left(k_{n}b_{n}+\oint_{S_{t}}\mathbf{H}\times\mathbf{E}_{n}\cdot d\mathbf{s}\right)\mathbf{H}_{n}\\
+\sum_{\alpha}\left(\oint_{S_{t}}\mathbf{H}\times\mathbf{F}_{\alpha}\cdot d\mathbf{s}\right)\mathbf{F}_{\alpha}\label{eq:curl_h}
\end{multline}

Both the solenoidal and irrotation modes are orthogonal \cite{collinbookfoundation},
and their amplitudes are chosen to satisfy the normalization condition,
i.e.,
\[
\langle\mathbf{E}_{n},\mathbf{E}_{n}\rangle_{V_{t}}=\langle\mathbf{F}_{\alpha},\mathbf{F}_{\alpha}\rangle_{V_{t}}=\langle\mathbf{H}_{n},\mathbf{H}_{n}\rangle_{V_{t}}=\langle\mathbf{G}_{\lambda},\mathbf{G}_{\lambda}\rangle_{V_{t}}=1,
\]
where the term $\langle\mathbf{X},\mathbf{Y}\rangle_{V_{t}}$ denote
the volumetric projection (inner product in real three-dimensional
vector space) of the two vector functions $\mathbf{X}$ and $\mathbf{Y}$
and is given by,
\[
\langle\mathbf{X},\mathbf{Y}\rangle_{V_{t}}=\int_{V_{t}}\mathbf{X}\cdot\mathbf{Y}dv,
\]
It is worth mentioning that the expansions in (\ref{eq:e_cav}), (\ref{eq:h_cav}),
(\ref{eq:curl_e}), and (\ref{eq:curl_h}) are strictly valid in the
volume integral sense, i.e. the equalities are satisfied for projections
(volume integrals) over arbitrary well-behaved vector functions.

\subsection{Problem Model and Boundary Conditions}

Referring to the general problem of Fig. \ref{Config}(a), the domain
of volume $V_{t}$ with known eigenmodes (Fig. \ref{Config}(b)) is
regarded as a fictitious cavity enclosing the actual cavity $V_{c}$,
which in turn may have a number of waveguide ports partially overlapping
with the fictitious cavity surface $S_{t}$ as seen in Fig. \ref{Config}(c).
In general the treatment that will be proposed can handle cavity partially
loaded with magnetic and dielectric materials. However, for the sake
of simplicity, the model will be formulated only for dielectric material
loading with relative permittivity $\varepsilon_{r}\left(\mathbf{r}\right)$.
Beside the loading dielectric material $\varepsilon_{r}$ inside the
volume $V_{c}$, the extra added space $\delta V$ (gray area in Fig.
\ref{Config}(c)) inside the canonical cavity is considered as a material
with very high conductivity $\sigma$. Hence, the problem in Fig.
\ref{Config}(a) reduces to a canonical cavity loaded with a dielectric
material, and the added highly conducting material that fills the
space $\delta V$, in addition to the ports.

The canonical cavity walls are considered either as PE, PM, or a combination
of both. In order to minimize the number of terms in the fields expansion,
it is preferable to choose the fictitious cavity $V_{t}$ to conform
as much as possible with the actual cavity $V_{c}$, hence minimizing
the added conducting material in region $\delta V$. Therefore most
of the boundaries are taken as PE, specially on boundaries coinciding
with those of the actual cavity $V_{c}$. However, some boundaries
may take as PM, specially those having waveguide ports through them.
The ports are the excitation apertures with boundaries that are generally
neither PE nor PM. But in forming the canonical cavity $V_{t}$, 
to make a convenient choice for the port boundaries (PE or PM) such
that the cavity solenoidal and irrotational electric and magnetic
eigenmodes can be analytically calculated easily. Boundary conditions
for the different modes are given in the Appendix in (\ref{eq:PM_BCs})
and (\ref{eq:PE_BCs}), where
they were used in calculating some of the modes for cylindrical and
rectangular cavities employed in this paper. Either the tangential
actual electric field or magnetic field vanishes on the canonical
cavity surface $S_{t}$ except on surfaces of the ports $S_{p}$. Therefore, the integrals in (\ref{eq:curl_e}) and (\ref{eq:curl_h})
on the closed surface $S_{t}$ vanish except on the port surfaces
$S_{p}$, i.e., $\oint_{S_{t}}\mathbf{E}\times\mathbf{H}_{n}\cdot d\mathbf{s}=\int_{S_{p}}\mathbf{E}\times\mathbf{H}_{n}\cdot d\mathbf{s}$.

Since these eigenmodes form a complete set, arbitrary field distributions
within $V_{t}$ can be expressed by (\ref{eq:e_cav}) and (\ref{eq:h_cav})
under arbitrary excitation (incident fields on ports) and loading
(inhomogeneous dielectric inclusions) conditions, as well as arbitrary
cavity shapes $V_{c}$ as long as they are a subset of $V_{t}$.

\subsection{Maxwell's Equations}

Substituting the expansions (\ref{eq:h_cav}) and (\ref{eq:curl_e})
into Maxwell's equation $\nabla\times\mathbf{E}=-j\omega\mu_{0}\mathbf{H}$
and performing eigenmode projection with $\mathbf{H}_{n}$ and $\mathbf{G}_{\lambda}$
and using the fact that the modes are orthonormal, yields \cite{slatterbokk,kurokawacavity,collinbookfoundation},
\begin{equation}
k_{n}a_{n}+\int_{S_{p}}\left(\mathbf{E}\times\mathbf{H}_{n}\right)\cdot d\mathbf{s}=-j\omega\mu_{0}b_{n}\label{eq:curl_E_H}
\end{equation}
\begin{equation}
\int_{S_{p}}\left(\mathbf{E}\times\mathbf{G}_{\lambda}\right)\cdot d\mathbf{s}=-j\omega\mu_{0}g_{\lambda}.\label{eq:curl_E_G}
\end{equation}
Similarly using the expansions (\ref{eq:e_cav}) and (\ref{eq:curl_h})
into Maxwell's equation $\nabla\times\mathbf{H}=j\omega\varepsilon_{0}\varepsilon_{r}(\mathbf{r})\mathbf{E}+\sigma(\mathbf{r})\mathbf{E}$
and performing eigenmode projection with $\mathbf{E}_{n}$ and $\mathbf{F}_{\alpha}$
yields,
\begin{multline}
k_{n}b_{n}+\int_{S_{p}}\left(\mathbf{H}\times\mathbf{E}_{n}\right)\cdot d\mathbf{s}=j\omega\varepsilon_{0}\times\\
\left[\sum_{n'}a_{n'}\left\langle \varepsilon_{rc}\mathbf{E}_{n},\mathbf{E}_{n'}\right\rangle _{V}+\sum_{\alpha'}f_{\alpha'}\left\langle \varepsilon_{rc}\mathbf{E}_{n},\mathbf{F}_{\alpha'}\right\rangle _{V}\right]\label{eq:curl_H_E}
\end{multline}
\begin{multline}
\int_{S_{p}}\left(\mathbf{H}\times\mathbf{F}_{\alpha}\right)\cdot d\mathbf{s}=j\omega\varepsilon_{0}\times\\
\left[\sum_{n'}a_{n'}\left\langle \varepsilon_{rc}\mathbf{F}_{\alpha},\mathbf{E}_{n'}\right\rangle _{V}+\sum_{\alpha'}f_{\alpha'}\left\langle \varepsilon_{rc}\mathbf{F}_{\alpha},\mathbf{F}_{\alpha'}\right\rangle _{V}\right]\label{eq:curl_H_F}
\end{multline}
where $\varepsilon_{rc}(\mathbf{r})=\varepsilon_{r}(\mathbf{r})-j\sigma(\mathbf{r})/\left(\omega\varepsilon_{0}\right)=\varepsilon_{r}(\mathbf{r})\left(1-j\tan\delta(\mathbf{r})\right)$
is the complex relative permitivity, which is a function of space
$\mathbf{r}\in V_{t}$, written in terms of the material loss tangent.
For low-loss dielectric material loading the cavity $\tan\delta\ll1$,
however for the added conducting medium in the added region $\delta V$,
$\tan\delta\gg1$, typically taken in the rest of the paper as $\tan\delta\sim10^{4}$.

Notice that the the two Maxwell's divergence equations $\nabla\cdot\mathbf{D}=\rho$
and $\nabla\cdot\mathbf{B}=0$ were not considered, as they can be
derived from the curl equations, together with charge continuity.
In fact, after some mathematical elaboration, (\ref{eq:curl_E_G})
can be obtained from $\nabla\cdot\mathbf{B}=0$, whereas (\ref{eq:curl_H_F})
can be derived from $\nabla\cdot\mathbf{D}=\rho$. Therefore equations
(\ref{eq:curl_E_H}) through (\ref{eq:curl_H_F}) represent the solution
of Maxwell's equation in terms of the coefficients $a_{n}$, $b_{n}$,
$f_{\alpha}$, and $g_{\lambda}$.

\section{Electromagnetic Cavity Resonance Analysis}

\begin{figure}
\centering
\includegraphics[scale=0.4]{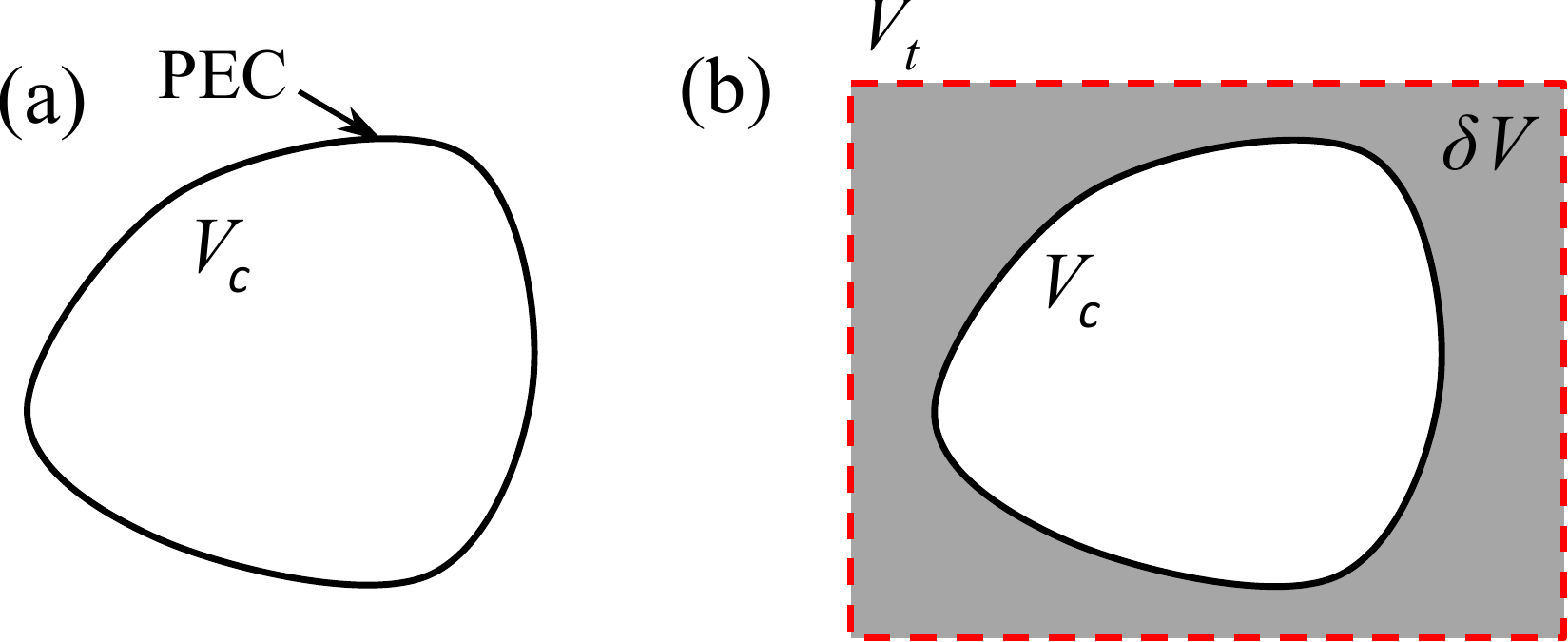}
\caption{\label{fig:Arbitrary-shaped-PEC-cavity.-1}(a) A conducting arbitrary-shaped
cavity. (b) Employed model of the cavity as part of a conducting canonical
cavity (dashed outline) partially filled with a highly conductive
material (grey colored regions).}
\end{figure}

A special case of the scenario depicted in Fig. \ref{Config}(a) is
that of a cavity resonance problem (Fig. \ref{fig:Arbitrary-shaped-PEC-cavity.-1}(a)),
viz., the cavity $V_{c}$ is not excited by external waveguides. This
problem was tackled in \cite{myAPS}, where the EPT was invoked in
a procedure which involved replacing the perfectly conducting cavity
surface by an equivalent surface current, expressed in terms of the
fictitious canonical cavity eigenmodes. The solution resulted in a
set of eigenvalues (resonance frequencies), some of which, however,
were not valid resonance modes, and mostly correspond to non-zero
fields in the added conducting region $\delta V$, hence named spurious
modes. Identifying and separating these spurious modes required exhaustive
post-processing, which was not successful in all cases in distinguishing
them.

A more efficient approach is proposed here wherein a highly (but not
perfectly) conducting material is used to fill in the residual part
of the canonical cavity $\delta V$ (the difference between the actual
cavity and canonical cavity volumes) as illustrated in Fig. \ref{fig:Arbitrary-shaped-PEC-cavity.-1}(b),
where it is simply considered as a material with very large loss tangent.
It is expected that the eigenvalues (resonance frequencies) in this
case will be complex and the spurious modes will be distinguished
upon comparing the real and imaginary parts of the eigenvalue (or
equivalently the quality factor $Q$) of the resulted modes.

\subsection{Air-Filled Arbitrary-Shaped Cavity}

Referring to Fig. \ref{fig:Arbitrary-shaped-PEC-cavity.-1}(a), and
assuming an air-filled cavity enclosed by a fictitious canonical cavity
with the difference volume filled with a highly conductive material
with a very high loss tangent $\tan\delta$ as in Fig. \ref{fig:Arbitrary-shaped-PEC-cavity.-1}(b),
the surface integrals representing coupling with port modes in (\ref{eq:curl_E_H})
through (\ref{eq:curl_H_F}) vanish. Hence, (\ref{eq:curl_E_G})
results in $g_{\lambda}=0$, and after substituting with $b_{n}$
from (\ref{eq:curl_E_H}) into (\ref{eq:curl_H_E}), a set of equations
in $a_{n}$ and $f_{\alpha}$ coefficients remains:
\begin{multline}
k_{n}^{2}a_{n}=k_{0}^{2}\left\{ \sum_{n'}a_{n'}\left(\delta_{n'n}-j\tan\delta\left\langle \mathbf{E}_{n},\mathbf{E}_{n'}\right\rangle _{\delta V}\right)\right.\\
\left.-j\tan\delta\sum_{\alpha'}f_{\alpha'}\left\langle \mathbf{E}_{n},\mathbf{F}_{\alpha'}\right\rangle _{\delta V}\right\} \label{eq:a_n_f_alpha_1}
\end{multline}
\begin{multline}
\sum_{\alpha'}f_{\alpha'}\left(\delta_{\alpha'\alpha}-j\tan\delta\left\langle \mathbf{F}_{\alpha},\mathbf{F}_{\alpha'}\right\rangle _{\delta V}\right)=\\
-j\tan\delta\sum_{n'}a_{n'}\left\langle \mathbf{F}_{\alpha},\mathbf{E}_{n'}\right\rangle _{\delta V}\label{eq:a_n_f_alpha_2}
\end{multline}
where $\langle\cdot,\cdot\rangle_{\delta V}$ denote the volume inner
product over the added conducting medium $\delta V$. Notice that
(\ref{eq:a_n_f_alpha_1}) and (\ref{eq:a_n_f_alpha_2}) were simplified
using the fact that modes are orthonormal, and there is no dielectric
material inside the cavity, except for the added conducting one in
the volume $\delta V$ which has $\varepsilon_{rc}=1-j\tan\delta$.
Combining (\ref{eq:a_n_f_alpha_1}) and (\ref{eq:a_n_f_alpha_2})
and writing them in matrix form, the problem can be cast in the form
of the eigenvalue problem
\begin{multline}
\frac{1}{k^{2}}\left[a\right]_{N\times1}=\left[K\right]^{-1}\left(\left[U\right]_{N\times N}-j\tan\delta\left[Q^{EE}\right]_{N\times N}+\right.\\
\left.j\tan\delta\left[Q^{EF}\right]_{N\times M}\left[Z\right]_{M\times M}\left[Q^{EF}\right]_{M\times N}^{T}\right)\left[a\right]_{N\times1},\label{eq:eigenvalue}
\end{multline}
\[
\left[Z\right]_{M\times N}=\left[\left[Q^{FF}\right]_{M\times M}-\frac{1}{j\tan\delta}\left[U\right]_{M\times M}\right]^{-1}
\]
where $\left[U\right]$ is the identity matrix, $k=2\pi f_{res}/c$
is the complex resonance wavenumber, $\left[a\right]_{N\times1}$
is a vector of the cavity field coefficients $a_{n}$ , and the matrices
$[Q^{EE}]$, $[Q^{EF}]$, and $[Q^{FF}]$ hold the values of the inner
products $\langle\mathbf{E}_{n},\mathbf{E}_{n'}\rangle_{\delta V}$,
$\langle\mathbf{E}_{n},\mathbf{F}_{\alpha'}\rangle_{\delta V}$, and
$\langle\mathbf{F}_{\alpha},\mathbf{F}_{\alpha'}\rangle_{\delta V}$,
respectively, whereas $\left[K\right]^{-1}$ is inverse of a diagonal
matrix $\left[K\right]$ with the diagonal elements $k_{n}^{2}$.

The eigenvalue problem in (\ref{eq:eigenvalue}) is solved for the
complex eigenvalues $1/k^{2}$. It is worth mentioning that although
practical conductors loss tangent is frequency dependent, it is assumed
that the added conductor in $\delta V$ has a constant very large
loss tangent over the frequency range of interest, which would be
almost identical with the practical case at microwave frequencies.

\subsection{Results and Convergence Study}

\begin{figure}
\centering
\includegraphics[scale=0.45]{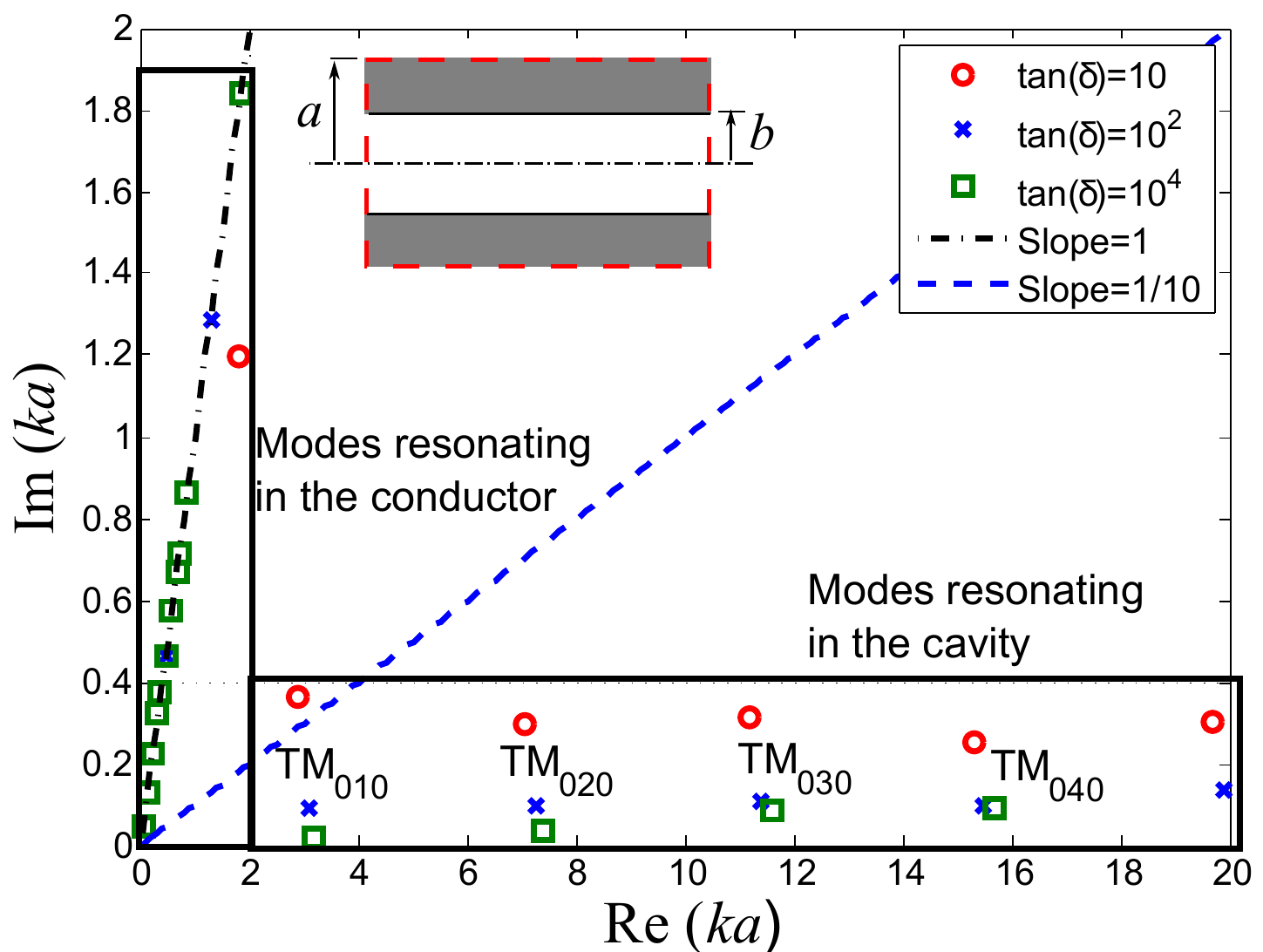}
\caption{\label{fig:cyl_cav_k} Complex mode wavenumbers for circular cylinder
case shown in the inset with $b$=1.5 cm, $a$=2 cm.}
\end{figure}

The previous approach is first verified for the canonical case of
a circular cylindrical cavity shown in inset of Fig. \ref{fig:cyl_cav_k},
i.e., the solution of the (actual) circular cavity of radius $b$
is obtained using the eigenmodes of another (fictitious) circular
cavity of radius $a$, where $a>b$. The symmetry of the structure
shown in the inset of Fig. \ref{fig:cyl_cav_k} (a body-of-revolution) permits us to deal
with axial symmetric modes, and implies that the axial dependence
of the fields in the actual cavity will be the same that in
the canonical one. Hence the axial dependence will be intentionally
ignored, focusing only on transverse radial variation. Fig. \ref{fig:cyl_cav_k}
plots the complex eigenvalue $k$ of (\ref{eq:eigenvalue}) for a
case where $b=1.5$ cm and $a=2$ cm, using $30$ eigenmodes of the
fictitious cavity.

It is obvious from the results that the eigenvalues are directly separated
into two sets: modes resonating in the conductor and modes resonating
inside the cavity. The former set corresponds to spurious modes, while
the latter set corresponds to the actual cavity modes. Thus, the mode
separation follows in a straight-forward manner, simply by comparing
the real and imaginary parts of the complex wavenumber. The actual
modes are those with high real part compared to the imaginary part
for high values of conductivity (i.e., high $Q$ factor). A rule-of-thumb
would be that $k_{r}/k_{i}>10$ for a mode to be considered physical.

The effect of increasing the conductivity on the accuracy of the first
resonance frequency is illustrated in Table \ref{tab:cond} keeping
the same number of eigenmodes ($N$=30) for the fictitious cavity.
Also as expected, the relative error in the resonance frequency decreases
as the number of eigenmodes is increased for the first resonance frequency,
where for $N=$10, 20, and 30 the relative error is 4.5551\%, 0.7104\%
and 0.0606\% respectively. A final remark evident in Fig. \ref{fig:cyl_cav_k}
about the spurious resonance inside the added conductor is that their
real and imaginary frequency parts are almost equal, a common criteria
for resonances occurring inside highly conducting media.

\begin{table}
\caption{\label{tab:cond} The effect of conductivity on the accuracy of the
obtained resonance frequencies}
\centering{}%
\begin{tabular}{cllll}
\hline
 & Theoretical  & $\tan\delta=10$  & $\tan\delta=10^{2}$  & $\tan\delta=10^{4}$\tabularnewline
\hline
$f_{\textrm{res}}$ GHz  & 7.6548  & 6.936  & 7.4194  & 7.6502 \tabularnewline
Error \%  & -  & 9.3894 \%  & 3.0749 \%  & 0.0606 \% \tabularnewline
\hline
\end{tabular}
\end{table}

\begin{figure}
\centering
\includegraphics[scale=0.45]{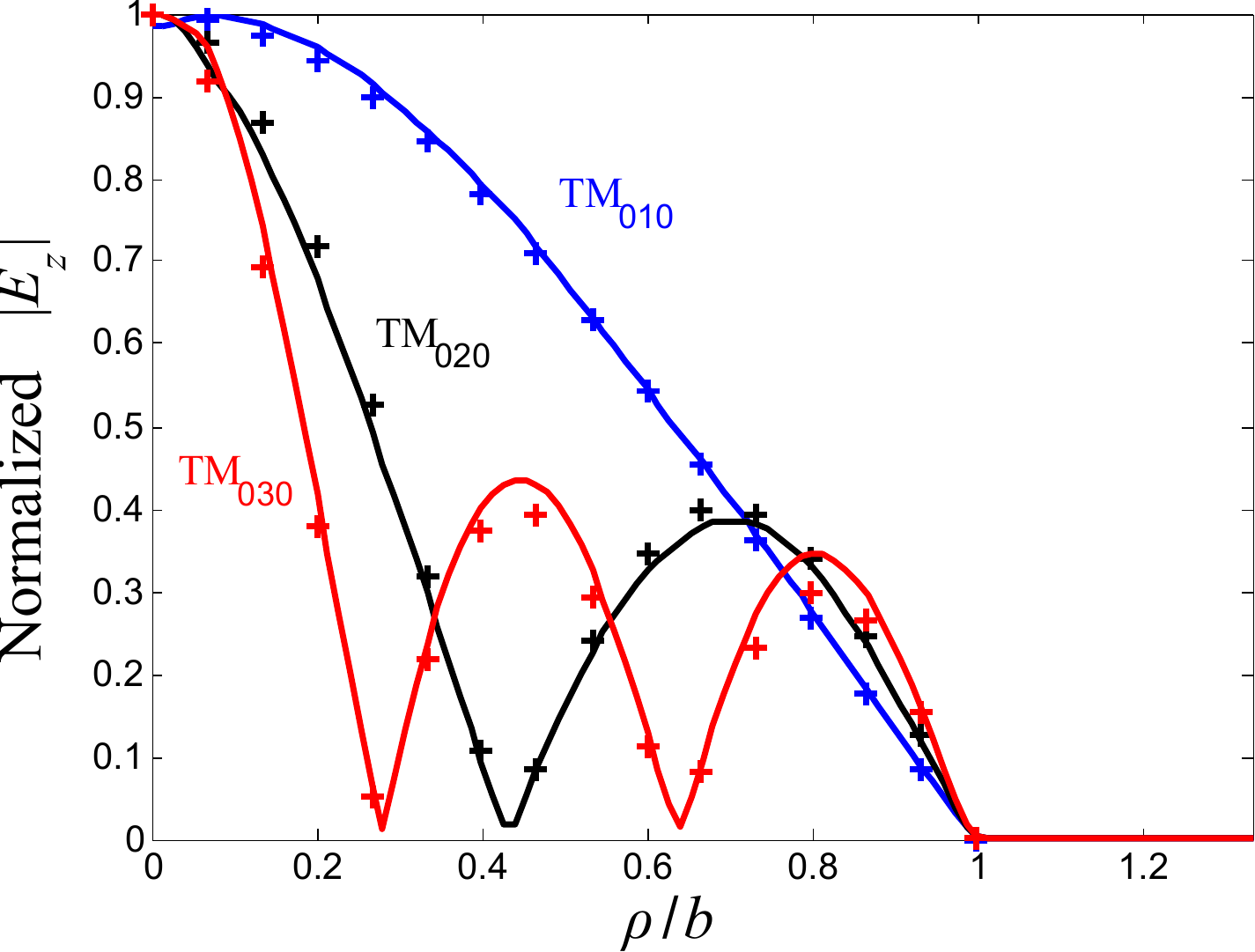}
\caption{\label{fig:cyl_cav_1st}Field plot versus radial distance for $\mathrm{TM_{010}}$,
$\mathrm{TM_{020}}$, and $\mathrm{TM_{030}}$ modes of cavity of radius
b=1.5 enclosed by canonical cavity of radius $a$ =2 cm. Crosses indicate
the EPT results for $N=40$, whereas the solid line indicates the exact ones.}
\end{figure}

The solution of the eigenvalue problem results also in the eigenvectors,
which represent the field coefficients of the eigenmode expansion
in (\ref{eq:e_cav}) and (\ref{eq:h_cav}) for each resonance. Figure
\ref{fig:cyl_cav_1st} illustrates the field plot inside the cavity
using the eigenvectors (coefficients) of $\mathrm{TM_{010}}$, $\mathrm{TM_{020}}$,
and $\mathrm{TM_{030}}$ modes exhibiting very good agreement with
the analytical solution. It is obvious that the fields tend to zero
in the conducting filling where $a<\rho<b$ smoothly, accurately,
and without any oscillations. It should be emphasized, however, that
the eigenmode expansion is understood in the volume-integral sense,
not in a point-wise sense \cite{Mam_journal}. Thus the integrals
of the field distributions are in general expected to be accurate,
but this is not always the case for their point-wise evaluation. As
the resonance frequencies, in the above formulation, depend on the
field volume integrals, their values are also expected to be accurate
as reported in the previous cases.

\begin{figure}
\centering
\includegraphics[width=7cm]{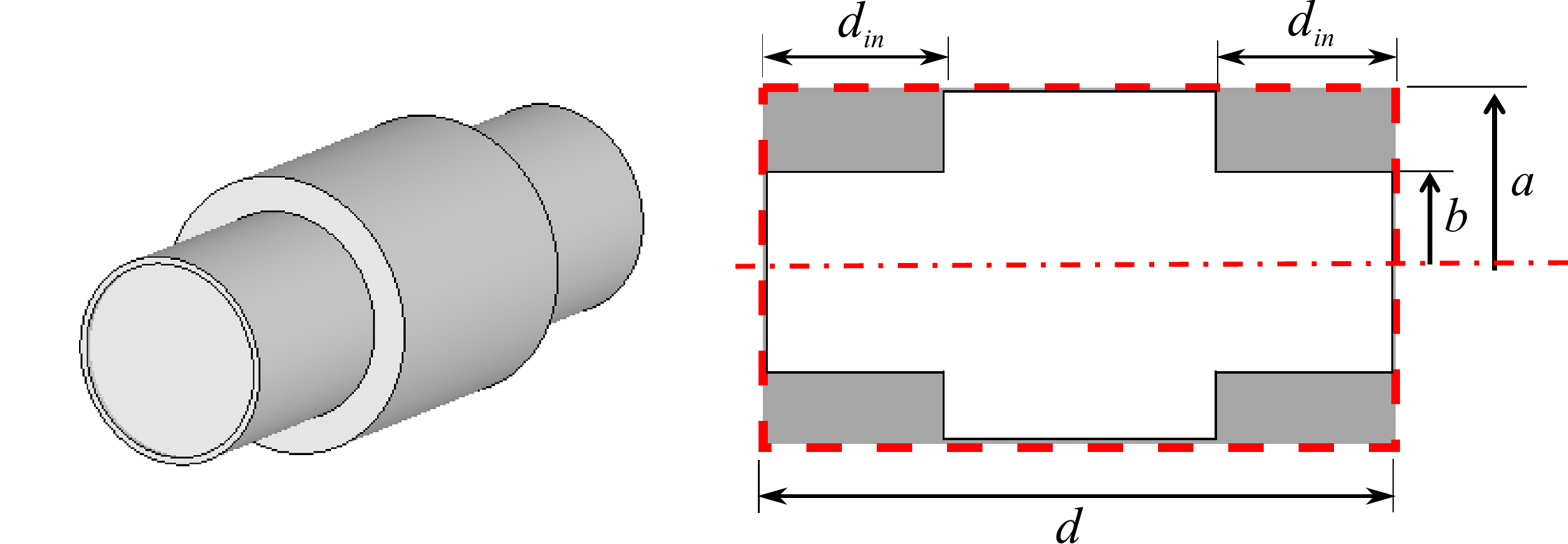}
\caption{\label{fig:stepped_cav}Stepped cavity enclosed by cylindrical canonical
cavity.}
\end{figure}

\begin{figure}
\centering
\includegraphics[scale=0.42]{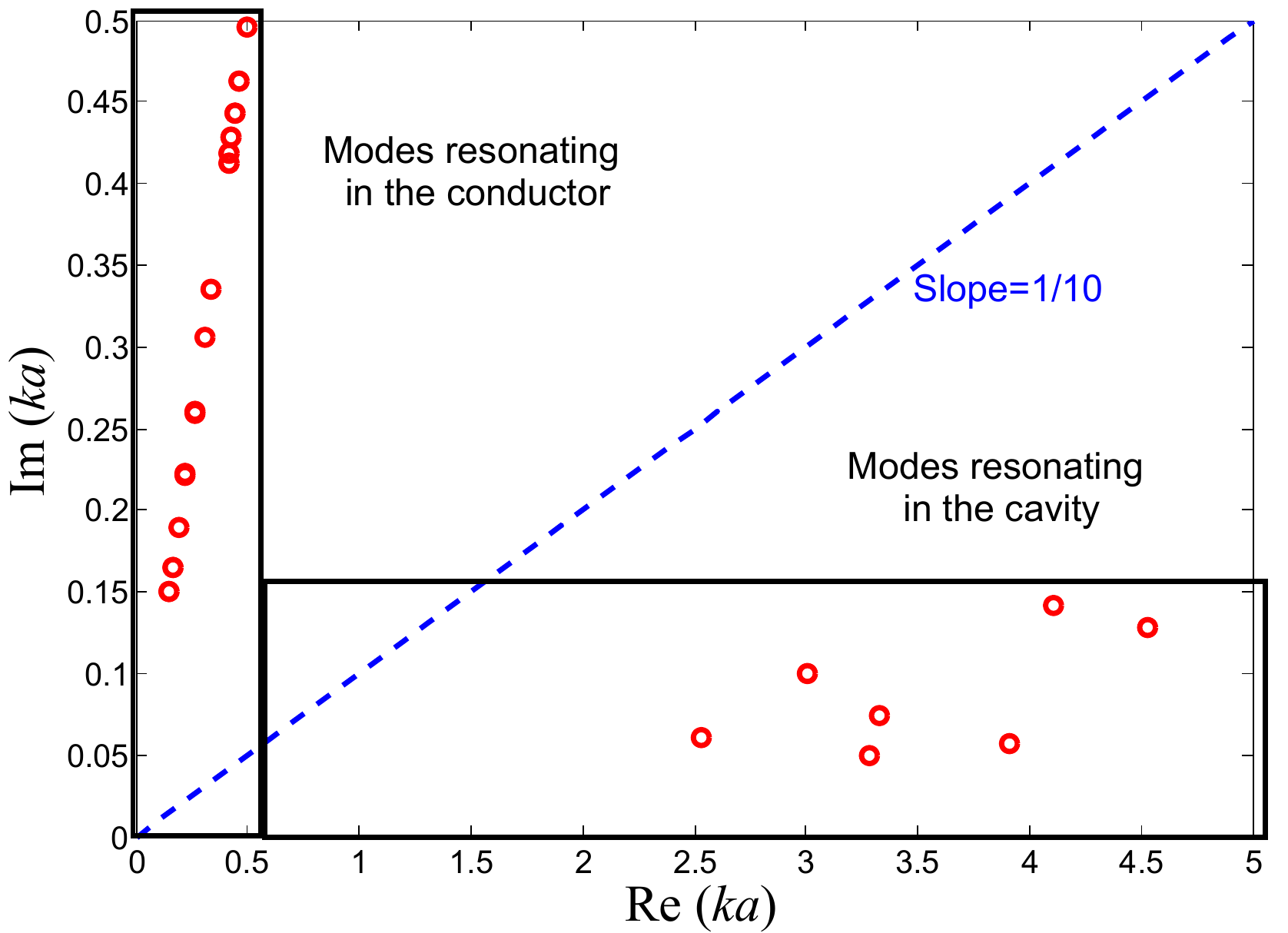}
\caption{\label{fig:stepped_cav_k}Complex wavenumbers for stepped case with
dimensions $a$=2 cm, $b$=1.5 cm, $d_{in}$=3 cm, and $d$=10 cm,
$N=400$ with equal number of expansion modes for each direction.}
\end{figure}

Second, the problem of a stepped cavity illustrated in Fig. \ref{fig:stepped_cav}
is considered, as an example of a homogeneous problem with sharp corners. The numerical solver
is used to obtain reference results for TM$^z$ resonating modes. The complex eigenvalues of stepped
cavity, with dimensions $a$=2 cm, $b$=1.5 cm, $d_{in}$=3 cm, and
$d$=10 cm, and $\tan\delta=10^{3}$ illustrated in Fig. \ref{fig:stepped_cav_k},
excellently converged to those obtained from full-wave simulations
(CST Microwave Studio based on the finite element method \cite{CST})
after applying the separation of spurious using the proposed rule-of-thumb
criteria, $k_{r}/k_{i}>10$, in a straightforward manner. Table \ref{tab:error_vs_cst} illustrates the small relative error
between the resonance frequencies obtained using CST and the EPT with
$N=400$ eigenmodes.

\begin{table}
\caption{\label{tab:error_vs_cst}The solution relative error of the first
three modes for stepped cavity with loss tangent $\tan\delta=10^{3}$. }
\centering{}%
\begin{tabular}{cccc}
\hline
\multicolumn{1}{c}{} & 1$^{\mathrm{st}}$Res  & 2$^{\mathrm{nd}}$Res  & 3$^{\mathrm{rd}}$Res\tabularnewline
\hline
CST  & 6.058 GHz  & 7.2076 GHz  & 7.9353 GHz\tabularnewline
\hline
EPT  & 6.0516 GHz  & 7.1805 GHz  & 7.8447 GHz\tabularnewline
\hline
\% Error  & 0.1056  & 0.376  & 1.1417\tabularnewline
\hline
\end{tabular}
\end{table}

\begin{figure}
\centering
\includegraphics[scale=0.35]{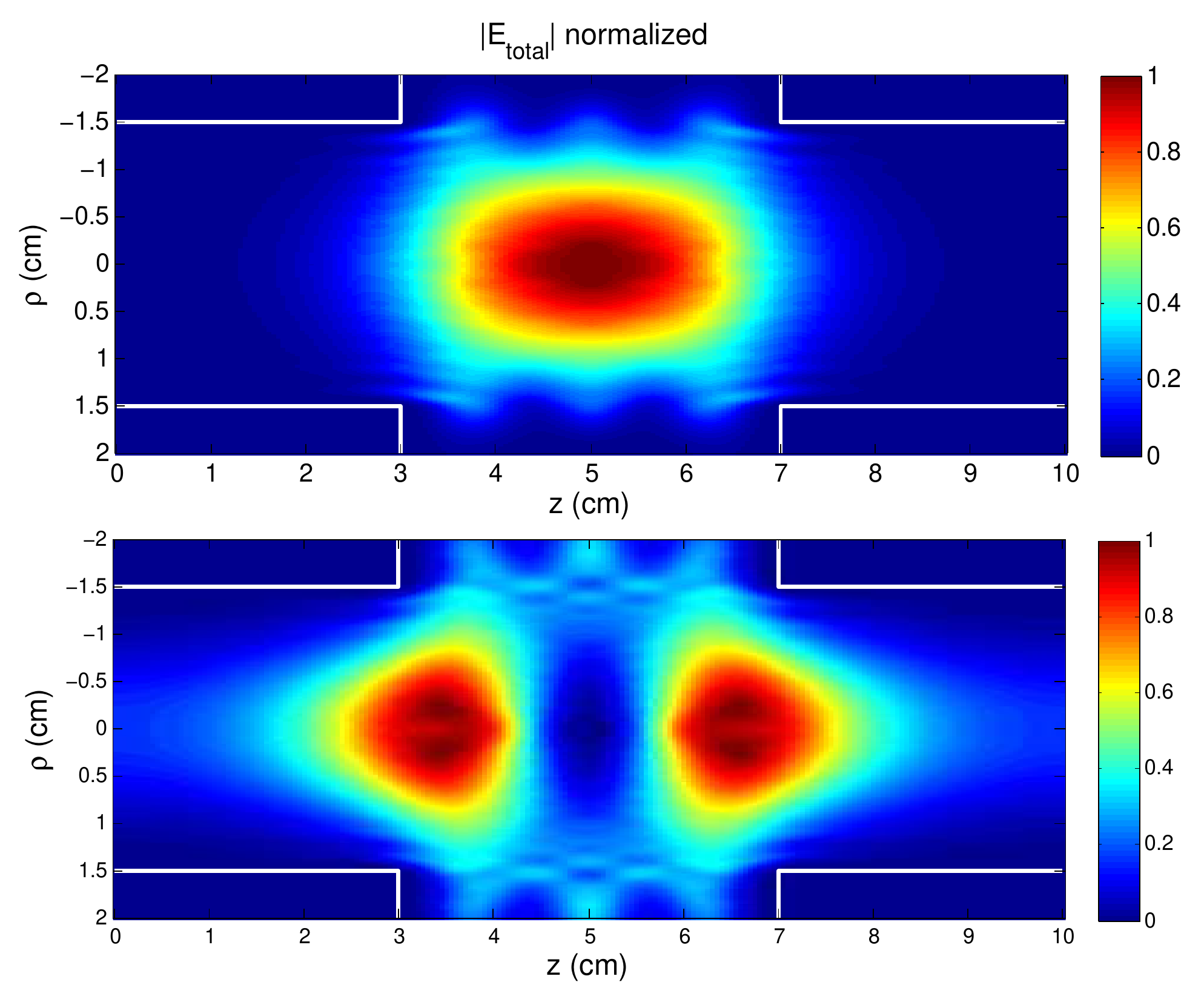}
\caption{\label{fig:stepped_1st}Two dimensional color map for the electric field distribution for the first two axially-symmetric
$\mathrm{TM^{z}}$ modes of cavity with dimensions $b$=1.5 cm, $d_{in}$=3
cm, and $d$=10 cm enclosed by canonical cavity of radius $a$=2 cm.}
\end{figure}

Figure \ref{fig:stepped_1st} shows the two dimensional color map
for the electric field distribution for the first
two axially-symmetric $\mathrm{TM^{z}}$ modes resonating inside the
cavity using the solution eigenvectors. The depicted modes exhibit
very good confinement within the physical cavity with almost vanishing
amplitude in the conducting residual parts of the fictitious canonical
cavity.

\begin{figure} [t]
\centering
\includegraphics[scale=0.4]{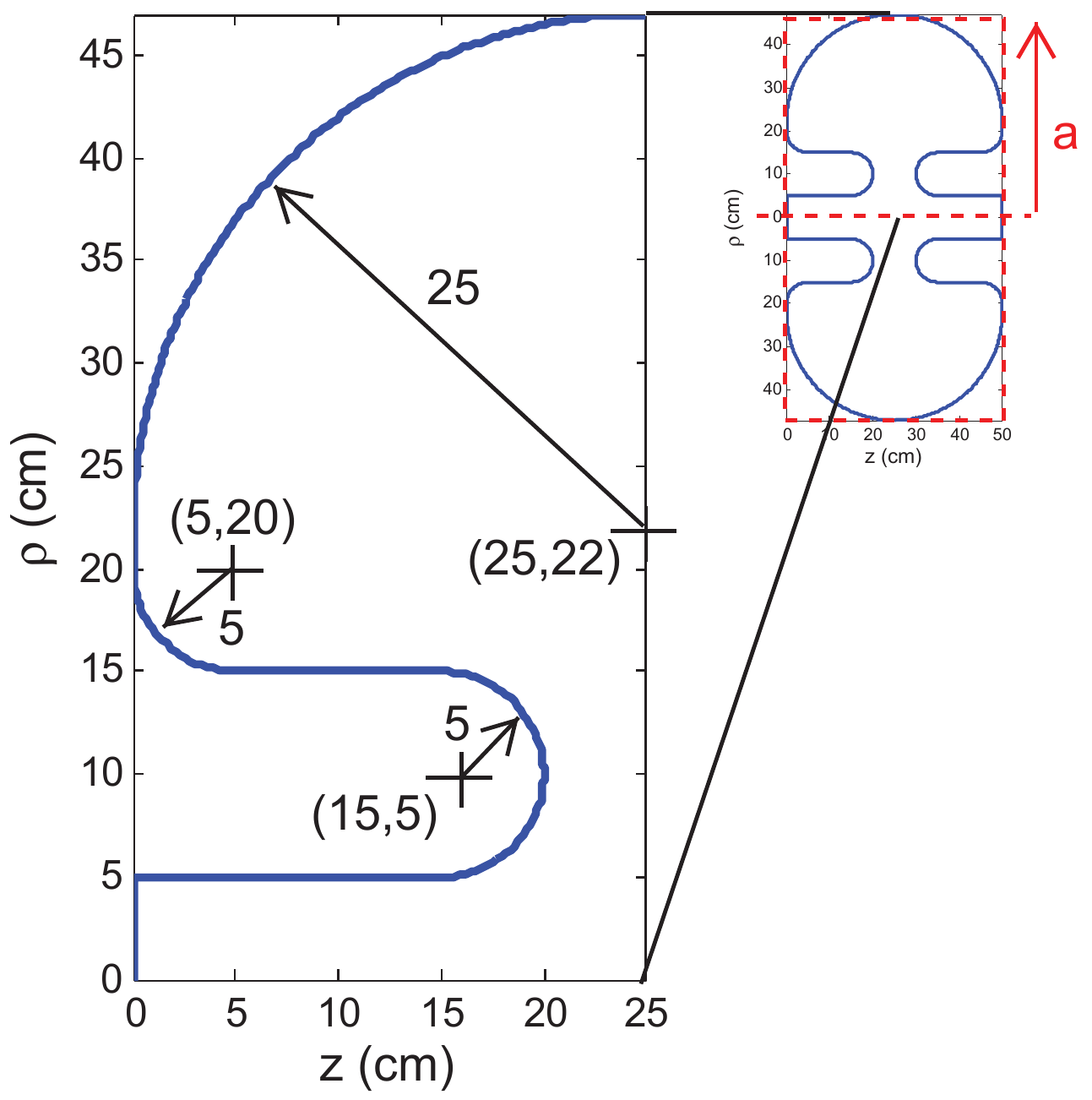}
\caption{\label{fig:linac_cav} Accelerating cavity with ``nose cone'' introduced
on each side of the accelerating gap, enclosed by a cylindrical canonical
cavity with radius of 47 cm and length of 50 cm.}
\end{figure}

Third, a more practical problem is considered with cavity shape that
is most common in standing wave linear accelerators (SW LINAC). In
order to increase the acceleration efficiency and concentrate the
electric field on the beam axis, a ‘nose cone’ is introduced on each
side of the accelerating gap \cite{linac} with a geometry illustrated
in Fig. \ref{fig:linac_cav}.

\begin{figure}
\centering
\includegraphics[scale=0.4]{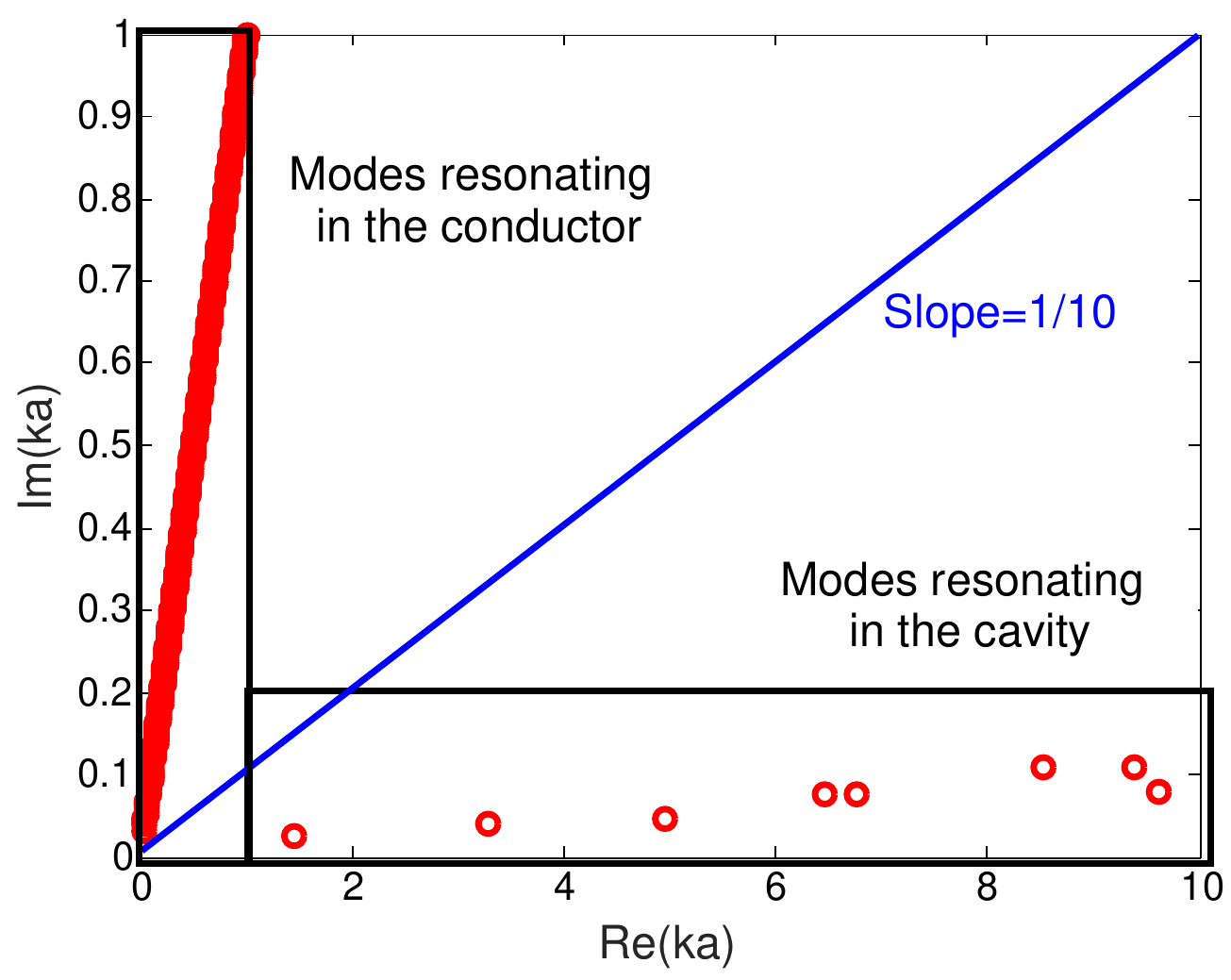}
\caption{\label{fig:linac_cav_k} Complex mode wavenumbers for the accelerator
cavity described in Fig. \ref{fig:linac_cav}.}
\end{figure}

\begin{figure} [t]
\centering
\includegraphics[scale=0.55]{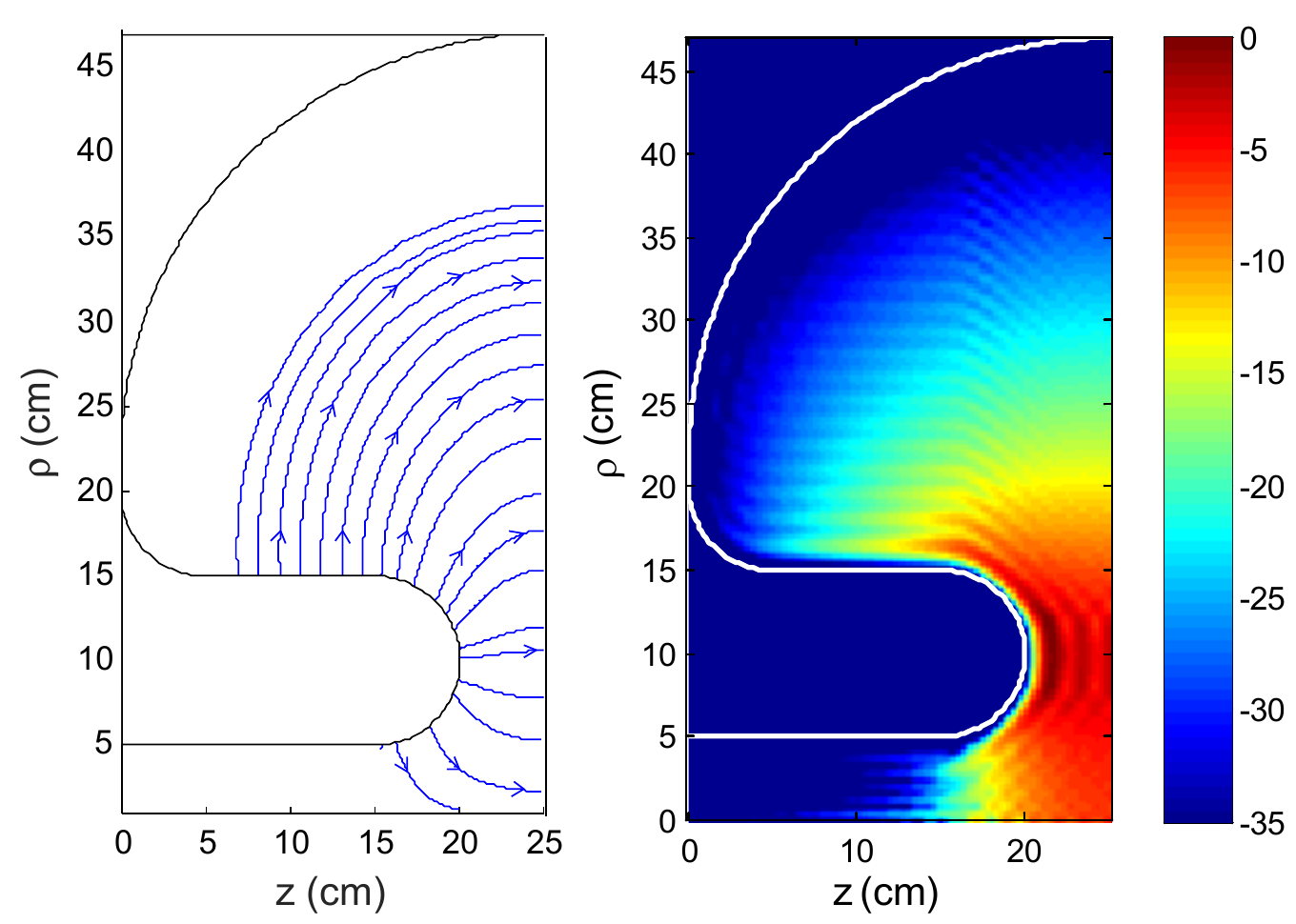}
\caption{\label{fig:linac_field_plot}Two dimensional field lines and color
map distribution of the electric field of the first
axially symmetric $\mathrm{TM^{z}}$ mode of cavity described in Fig.
\ref{fig:linac_cav}.}
\end{figure}

Figure \ref{fig:linac_cav_k} plots the complex eigenvalue solution
$k$, using $100$ canonical eigenmodes in each of the radial and
axial directions, and $\tan\delta=10^{4}$ for $\mathrm{TM^{z}}$
axially-symmetric modes. It can be observed that even with such a
relatively complex structure, the physical modes are directly separable
from spurious modes. The frequencies of the first two modes are 150
and 338 MHz which are only 5\% and 2.5\% off the results obtained
using full-wave simulation (CST), respectively.

Figure \ref{fig:linac_field_plot} shows the two dimensional color
map distribution of the field as well as the field lines plot for
the first axially symmetric $\mathrm{TM^{z}}$ mode resonating inside
the cavity using the solution eigenvectors. The results confirms the
previously mentioned effect of the `nose cone' in increasing the
field concentration on the beam axis. Also, the mode exhibits very
good confinement within the physical cavity with almost vanishing
amplitude in the conducting residual parts of the fictitious canonical
cavity as in previous cases.

\subsection{Extension to Dielectric-Loaded Cavities}

\begin{figure}
\centering
\begin{subfigure}{.5\textwidth}
\centering
\includegraphics[scale=0.4]{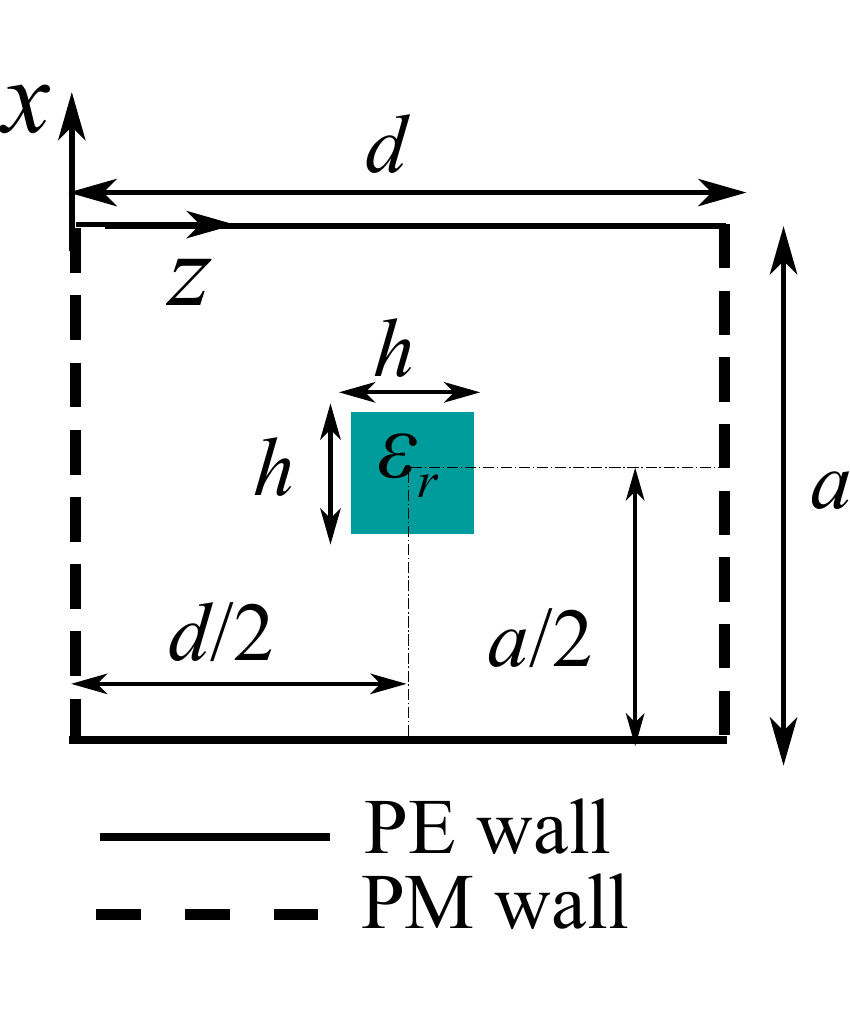}
\caption{}
\end{subfigure}
\\
\begin{subfigure}{.5\textwidth}
\centering
\includegraphics[scale=0.33]{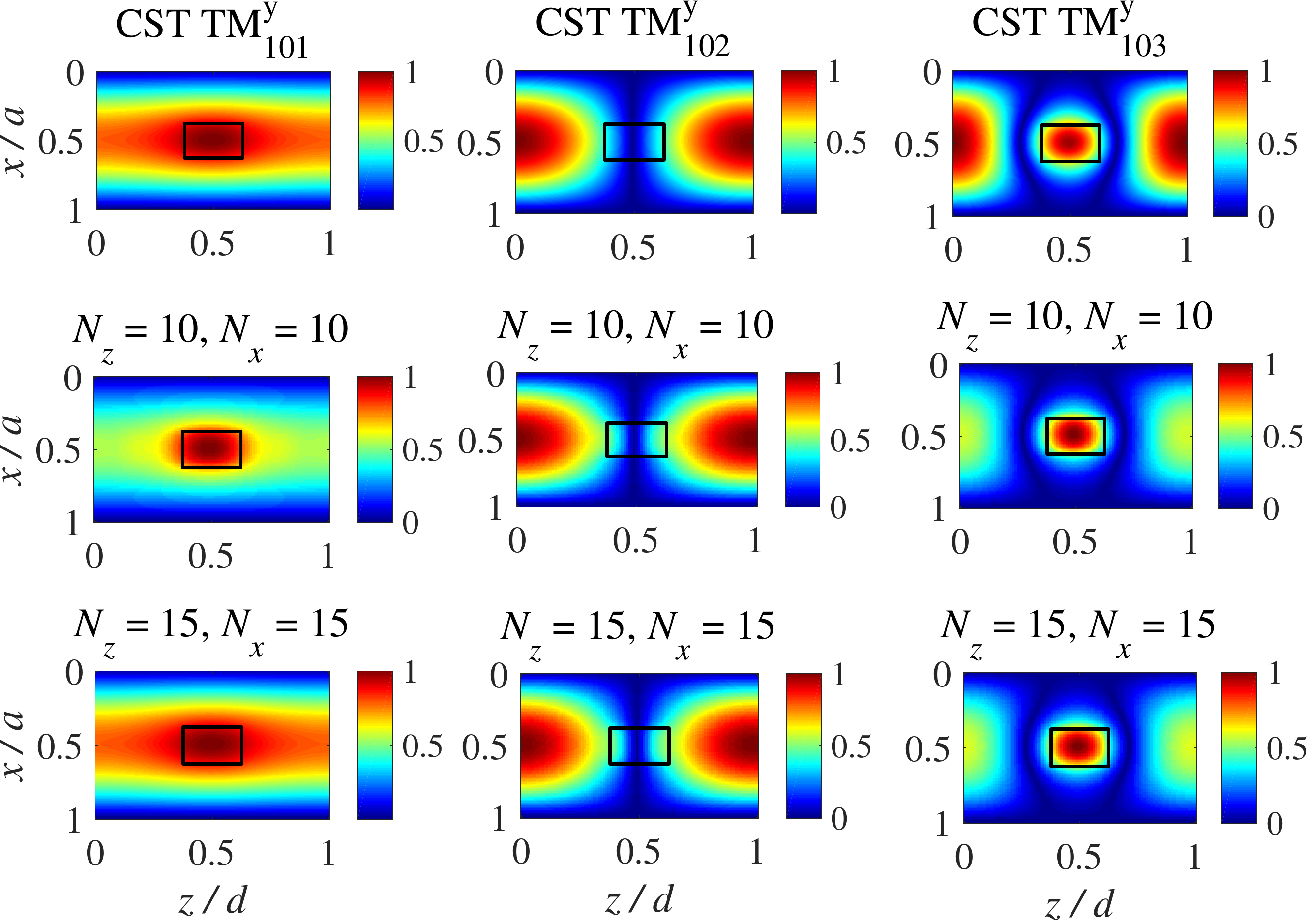}
\caption{}
\end{subfigure}

\caption{\label{fig_dii}(a) A rectangular cavity loaded with a dielectric
square obstacle, with mixed PE/PM boundary conditions. (b) Normalized
electric field $|E_{y}|$ map of the first three resonant modes obtained using
CST (first row) and using the EPT for increasing $N=N_x\times N_z$ eigenmodes. The square black line denotes
the location of the dielectric post.}
\end{figure}

The proposed EPT can be adapted in a straightforward manner to handle
cavities arbitrarily loaded with dielectric materials. Consider the case shown in Fig. \ref{fig_dii}(a), where a rectangular cavity resonator loaded with a dielectric post is depicted. The resonant ${\rm TM}_{n_{1}0n_{2}}^{y}$ modes are analyzed for cavity dimensions of
20 cm $\times$ 20 cm and square dielectric post having $\varepsilon_{r}=3$ and side $h=0.5$ cm centered with respect to the cavity. Moreover, the cavity is assumed to have mixed PE/PM boundaries (the same example will be investigated later in Section IV for a waveguide scattering problem). Adaptation of the EPT to this problem is straightforward: the factor $1-j\tan\delta$ in the previous formulation will be replaced with the relative permittivity of the post. Accordingly, the resonance frequencies are obtained as discussed in Section III-A. In Table \ref{tab:Resonance-Frequencies-for},
the resonance frequencies are calculated for the first three modes
using the EPT, showing fast convergence, stability with excessive number of modes, and good agreement with results obtained using CST (error less than 0.1 \% for $N=150$). Further investigation of the convergence of the field distribution is reported in the color map of the first three modes
for different values of $N$ in Fig. \ref{fig_dii}(b) in the $z$ and $x$ directions. Good field convergence is achieved despite the presence
of the sharp dielectric corners for 15$\times$15 eigenmodes, noting that more eigenmodes are needed for the higher order ${\rm TM}_{103}^{y}$.

The previous results have demonstrated the versatility of the EPT
technique in solving any arbitrary cavity resonance problem efficiently
and in a straight forward manner using a semi-analytical approach verified for canonical problems as well as other practical
and general problems. Next, analysis of
waveguides loaded with dielectrics is investigated.

\section{Analysis of Waveguide Discontinuities}

Beside the EPT capabilities of solving cavity resonance problems,
the EPT lends itself to the solution of electromagnetic scattering
problems from arbitrary dielectric discontinuities inside conducting
waveguides \cite{othman_IMS2013}. In this section, scattering
from dielectric objects in straight and bent waveguide sections are analyzed.
For convenience, the inner product $\langle\cdot,\cdot\rangle_{S_{p}}$
will be used to denote integrals over the port surface $S_{p}$ in
this section, and is not to be confused with the volume integrals
in the previous section.

\begin{table}
\caption{\label{tab:Resonance-Frequencies-for}Resonance Frequencies for the
first three ${\rm TM}_{10n}^{y}$ modes of the rectangular cavity
shown in Fig. \ref{fig_dii}.}
\centering{}%
\begin{tabular}{|c|c|c|c|}
\cline{2-4}
\multicolumn{1}{c|}{} & ${\rm TM}_{103}^{y}$ & ${\rm TM}_{103}^{y}$ & ${\rm TM}_{103}^{y}$\tabularnewline
\hline
CST & 6.6465 & 10.4599 & 14.5548\tabularnewline
\hline
$N$=50 & 6.6595 & 10.4727 & 14.6023\tabularnewline
\hline
$N$=150 & 6.6513 & 10.4673 & 14.5661\tabularnewline
\hline
$N$=500 & 6.6511 & 10.4671 & 14.5648\tabularnewline
\hline
$N$=1000 & 6.6511 & 10.4672 & 14.5646\tabularnewline
\hline
\end{tabular}
\end{table}

\subsection{Fields on Waveguide Port}

The fields tangential to the ports surface $\mathbf{n}\times\mathbf{E}$
and $\mathbf{n}\times\mathbf{H}$ can be described in terms waveguide
modal expansions:
\begin{equation}
\mathbf{n}\times\mathbf{E}=\sum_{p}V_{p}\mathbf{n}\times\mathbf{e}_{p},\label{eq:n_cross_E_expansion}
\end{equation}
\begin{equation}
\mathbf{n}\times\mathbf{H}=\sum_{p}I_{p}Z_{p}\mathbf{n}\times\mathbf{h}_{p}=\sum_{p}I_{p}\mathbf{e}_{p},\label{eq:n_cross_H_expansion}
\end{equation}
where $\mathbf{e}_{p}$ and $\mathbf{h}_{p}$ are the reference waveguide
mode electric and magnetic fields, respectively, which are related
through,
\begin{equation}
\mathbf{n}\times\mathbf{h_p}=Y_{p}\mathbf{e}_{p},\label{eq:waveport_e_h}
\end{equation}
where $Y_{p}$ and $Z_{p}=1/Y_{p}$ are, respectively, the waveport
mode admittance and impedance. In order to uniquely define the amplitudes
$V_{p}$ and $I_{p}$, some normalization has to be adopted in terms
of $\mathbf{e}_{p}$ and/or $\mathbf{h}_{p}$. The port transverse
modal fields $\mathbf{e}_{p}$ and $\mathbf{h}_{p}$ (which correspond
to the excited TE and TM waveguide modes) must be carefully normalized
to take into consideration the evanescent modes, such that the GSM
formulation can handle asymmetric impedance matrices properly, as
in \cite{impednormhjaskal,wvgcttheroty,levyevancescmodes}. For simplicity,
it is assumed that the field $\mathbf{e}_{p}$ is real and is normalized
such that,
\begin{equation}
\langle\mathbf{e}_{p},\mathbf{e}_{p}\rangle_{S_{p}}=1.\label{eq:waveport_normalization}
\end{equation}
The admittance $Y_{p}$ of the $p^{\text{th}}$ waveguide port mode
is given by,
\begin{equation}
Y_{p}=\frac{1}{\eta_{0}}\begin{cases}
{\displaystyle \frac{1}{\sqrt{1-k_{c}^{2}/k_{0}^{2}}}} & \textrm{TE}\text{ prop., }k_{0}>k_{c}\\
{\displaystyle \frac{j}{\sqrt{k_{c}^{2}/k_{0}^{2}-1}}} & \textrm{TE}\text{ Evan., }k_{0}<k_{c}\\
{\displaystyle \sqrt{1-k_{c}^{2}/k_{0}^{2}}} & \textrm{TM}\text{ prop., }k_{0}>k_{c}\\
{\displaystyle -j\sqrt{k_{c}^{2}/k_{0}^{2}-1}} & \textrm{TM}\text{ Evan., }k_{0}<k_{c}
\end{cases}\label{eq:admittance_exp}
\end{equation}
where $k_{0}=\omega/c$ and $\eta_{0}$ are the free space wavenumber
and intrinsic impedance, respectively, and $k_{c}$
is the cutoff wavenumber of the external TE/TM waveguide modes.

In contrast with cavity resonance problems, for problems with waveguide
ports, not all surface port integrals that appear in (\ref{eq:curl_E_H})
through (\ref{eq:curl_H_F}) will vanish. To make use of the eigenmodes
of some canonical cavity, however, suitable boundary conditions must
be imposed on the ports. Here, PM boundaries will be assumed, i.e.,
$\mathbf{n}\times\mathbf{H}_{n}=\mathbf{n}\cdot\mathbf{E}_{n}=0$,
and $\partial\Phi_{\alpha}/\partial n=\Psi_{\lambda}=0$. Thus, the
vanishing of surface integrals in (\ref{eq:curl_E_H}) gives $k_{n}a_{n}=-j\omega\mu_{0}b_{n}$
and that in (\ref{eq:curl_E_G}) gives $g_{\lambda}=0$, similar to
the corresponding results for closed cavity case. Using these results,
the remaining port integrals are those containing the magnetic field
$\mathbf{H}$ in (\ref{eq:curl_H_E}) and (\ref{eq:curl_H_F}) become,
\begin{multline}
k_{n}^{2}a_{n}-jk_{0}\eta_{0}\sum_{p}I_{p}\langle\mathbf{e}_{p},\mathbf{E}_{n}\rangle_{S_{p}}=\\
k_{0}^{2}\left[\sum_{n'}a_{n'}\left\langle \varepsilon_{rc}\mathbf{E}_{n},\mathbf{E}_{n'}\right\rangle _{V_{t}}+\sum_{\alpha'}f_{\alpha'}\left\langle \varepsilon_{rc}\mathbf{E}_{n},\mathbf{F}_{\alpha'}\right\rangle _{V_{t}}\right]\label{eq:curl_H_E_port}
\end{multline}
where $\langle\cdot,\cdot\rangle_{S_{p}}$ is projection on the port
surface, i.e.,
\[
\langle\mathbf{e}_{p},\mathbf{E}_{n}\rangle_{S_{p}}=\int_{S_{p}}\mathbf{e}_{p}\cdot\mathbf{E}_{n}ds.
\]
Similarly the surface integral in (\ref{eq:curl_H_F}) can be manipulated
to give,
\begin{multline}
-j\eta_{0}\sum_{p}I_{p}\langle\mathbf{e}_{p},\mathbf{F}_{\alpha}\rangle_{S_{p}}=\\
k_{0}\left[\sum_{n'}a_{n'}\left\langle \varepsilon_{rc}\mathbf{F}_{\alpha},\mathbf{E}_{n'}\right\rangle _{V_{t}}+\sum_{\alpha'}f_{\alpha'}\left\langle \varepsilon_{rc}\mathbf{F}_{\alpha},\mathbf{F}_{\alpha'}\right\rangle _{V_{t}}\right]\label{eq:curl_H_F_port}
\end{multline}

Equations (\ref{eq:curl_H_E_port}) and (\ref{eq:curl_H_F_port})
can be cast in the following matrix form,
\begin{multline}
\left[\mathcal{G}\right]_{\left(M+N\right)\times\left(M+N\right)}\left[\begin{array}{c}
\left[a\right]_{N\times1}\\
\hdashline\left[f\right]_{M\times1}
\end{array}\right]=\\
\frac{-j\eta_{0}}{k_{0}}\left[\begin{array}{c}
\left[Q^{Ee}\right]_{N\times P}\\
\hdashline\left[Q^{Fe}\right]_{M\times P}
\end{array}\right]\left[I\right]_{P\times1}\label{eq:a_f_I_relation}
\end{multline}
where the super-matrix $\left[\mathcal{G}\right]$ is given by,
\begin{multline}
\left[\mathcal{G}\right]_{\left(N+M\right)\times\left(N+M\right)}=\\
\left[\begin{array}{cc}
\left[\varepsilon_{rc}Q^{EE}\right]_{N\times N}-\left[K\right]_{N\times N}/k_{0}^{2} & \left[\varepsilon_{rc}Q^{EF}\right]_{N\times M}\\
\hdashline\left[\varepsilon_{rc}Q^{EF}\right]_{M\times N}^{T} & \left[\varepsilon_{rc}Q^{FF}\right]_{M\times M}
\end{array}\right]\label{eq:G_matrix}
\end{multline}
where the sub-matrices shown in (\ref{eq:a_f_I_relation}) and (\ref{eq:G_matrix})
are described as follows:
\begin{description}
\item [{$\left[Q^{Ee}\right]_{N\times P}$}] \hfill{}Matrix with entries
$\langle\mathbf{E}_{n},\mathbf{e}_{p}\rangle_{S_{p}}$,
\item [{$\left[Q^{Fe}\right]_{M\times P}$}] \hfill{}Matrix with entries
$\langle\mathbf{F}_{\alpha},\mathbf{e}_{p}\rangle_{S_{p}}$,
\item [{$\left[\varepsilon_{rc}Q^{EE}\right]_{N\times N}$}] \hfill{}Matrix
with entries $\langle\varepsilon_{rc}\mathbf{E}_{n},\mathbf{E}_{n'}\rangle_{V_{t}}$,
\item [{$\left[\varepsilon_{rc}Q^{EF}\right]_{N\times M}$}] \hfill{}Matrix
with entries $\langle\varepsilon_{rc}\mathbf{E}_{n},\mathbf{F}_{\alpha}\rangle_{V_{t}}$,
\item [{$\left[\varepsilon_{rc}Q^{FF}\right]_{M\times M}$}] \hfill{}Matrix
with entries $\langle\varepsilon_{rc}\mathbf{F}_{\alpha'},\mathbf{F}_{\alpha}\rangle_{V_{t}}$.
\end{description}
Equation (\ref{eq:a_f_I_relation}) can be solved by inverting the
matrix $\mathcal{G}$ given by (\ref{eq:G_matrix}) to get the coefficients
$a_{n}$ and $f_{\alpha}$ in term of the ports excitation \textquotedbl{}current\textquotedbl{}
amplitudes $I_{p}$.

\subsection{Matching the Port and Cavity Fields}

To determine the scattered port fields, the next step would be to
match the external (port) and internal (cavity) fields, such that
\begin{equation}
\sum_{p}V_{p}\mathbf{n}\times\mathbf{e}_{p}=\mathbf{n}\times\mathbf{E},\qquad\text{on }S_{p}\label{19}
\end{equation}
By projecting both sides of the equation on the port modal field $\mathbf{h}_{p'}$,
one gets,
\begin{equation}
\sum_{p}V_{p}\langle\mathbf{n}\times\mathbf{e}_{p},\mathbf{h}_{p'}\rangle_{S_{p}}=\langle\mathbf{n}\times\mathbf{E},\mathbf{h}_{p'}\rangle_{S_{p}}\label{20}
\end{equation}
Due to the nature of the eigenmode expansions, the total cavity field
$\mathbf{E}$ may exhibit a discontinuity at the boundaries of the
expansion domain, viz. the cavity, even though this field is expanded
in terms of piecewise continuously differentiable eigenmodes within
$V_{t}$. This mathematical artifact brings about an overshoot in
the field at the boundary; a phenomenon that resembles Gibbs phenomenon
in Fourier series expansions \cite{gibbswebsuite}, for example. Consequently,
the point-wise matching of fields cannot be rigorously enforced. Therefore
the expansion (\ref{eq:e_cav}) cannot be strictly used except in
the volume integral forms, however after some manipulation \cite{Mam_journal},
it can be shown to be valid substitution in the surface integral in
Eq. (\ref{20}) under the condition of considering the port boundaries
as PM when calculating the canonical cavity solenoidal and irrotational
modes. When the waveport modes orthogonality and normalization condition
in (\ref{eq:waveport_normalization}) are applied on equation (\ref{20}),
it gives $V_{p}$ in terms of the coefficients $a_{n}$ and $f_{\alpha}$,
\begin{equation}
V_{p}=\sum_{n}a_{n}\langle\mathbf{e}_{p'},\mathbf{E}_{n}\rangle_{S_{p}}+\sum_{\alpha}f_{\alpha}\langle\mathbf{e}_{p'},\mathbf{F}_{\alpha}\rangle_{S_{p}}.\label{21}
\end{equation}
Solving equation (\ref{eq:a_f_I_relation}) for the vector containing
the coefficients $a_{n}$ and $f_{\alpha}$, and substituting in (\ref{21}),
we get the Z-matrix representation,
\begin{equation}
\left[V\right]_{P\times1}=\left[Z\right]_{P\times P}\left[I\right]_{P\times1},\label{eq:V_I_relation}
\end{equation}
where the impedance matrix $\left[Z\right]_{P\times P}$ is given
by, 
\begin{multline}
\left[Z\right]_{P\times P}=\frac{-j\eta_{0}}{k_{0}}\left[\begin{array}{cc}
\left[Q^{Ee}\right]_{P\times N}^{T} & \left[Q^{Fe}\right]_{P\times M}^{T}\end{array}\right]\times\\
\left[\mathcal{G}\right]_{\left(N+M\right)\times\left(N+M\right)}^{-1}\left[\begin{array}{c}
\left[Q^{Ee}\right]_{N\times P}\\
\hdashline\left[Q^{Fe}\right]_{M\times P}
\end{array}\right]\label{eq:Z}
\end{multline}
purely imaginary for lossless case where $\varepsilon_{rc}$ is real.

Now in order to get the scattering parameters $\left[S\right]$, the
following relation for the incident and scattered wave amplitudes
will be used.
\[
V_{p}=V_{p}^{+}+V_{p}^{-},\qquad I_{p}Z_{p}=V_{p}^{+}-V_{p}^{-}
\]
\[
\left[V_{p}^{-}\right]=\left[S\right]\left[V_{p}^{+}\right]
\]
\begin{equation}
[S]=\left(\left[Z\right]\left[\Upsilon\right]-\left[U\right]\right)\left(\left[Z\right]\left[\Upsilon\right]+\left[U\right]\right)^{-1},\label{eq:S}
\end{equation}
where $\left[\Upsilon\right]_{P\times P}$ is a diagonal matrix with
diagonal elements $Y_{p}$. The generalized scattering matrix GSM
can be obtained from the normalized impedance matrix $\left[\bar{Z}\right]$,
\[
\left[\bar{Z}\right]=\left[\sqrt{\Upsilon}\right]\left[Z\right]\left[\sqrt{\Upsilon}\right],
\]
\begin{equation}
\left[S^{g}\right]=\left(\left[\bar{Z}\right]-\left[U\right]\right)\left(\left[\bar{Z}\right]+\left[U\right]\right)^{-1}\label{GSM_Z}
\end{equation}
where $\left[\sqrt{\Upsilon}\right]_{P\times P}$ is a diagonal matrix
with diagonal elements $+\sqrt{Y_{p}}$ . It can be easily shown that
the relation between $\left[S\right]$ (\ref{eq:S}) and $\left[S^{g}\right]$
(\ref{GSM_Z}) is,
\begin{equation}
\left[S^{g}\right]=\left[\sqrt{\Upsilon}\right]\left[S\right]\left[\sqrt{\Upsilon}\right]^{-1}\label{GSM_S}
\end{equation}
It should be mentioned that there will ambiguity in calculating $\left[\sqrt{\Upsilon}\right]$
in (\ref{GSM_Z}) and (\ref{GSM_S}) due to the evanescent modes in the lossless waveguide.

\subsection{Dielectric Loaded Waveguides 2D Examples}

\begin{figure}
\centering{}\includegraphics[scale=0.45]{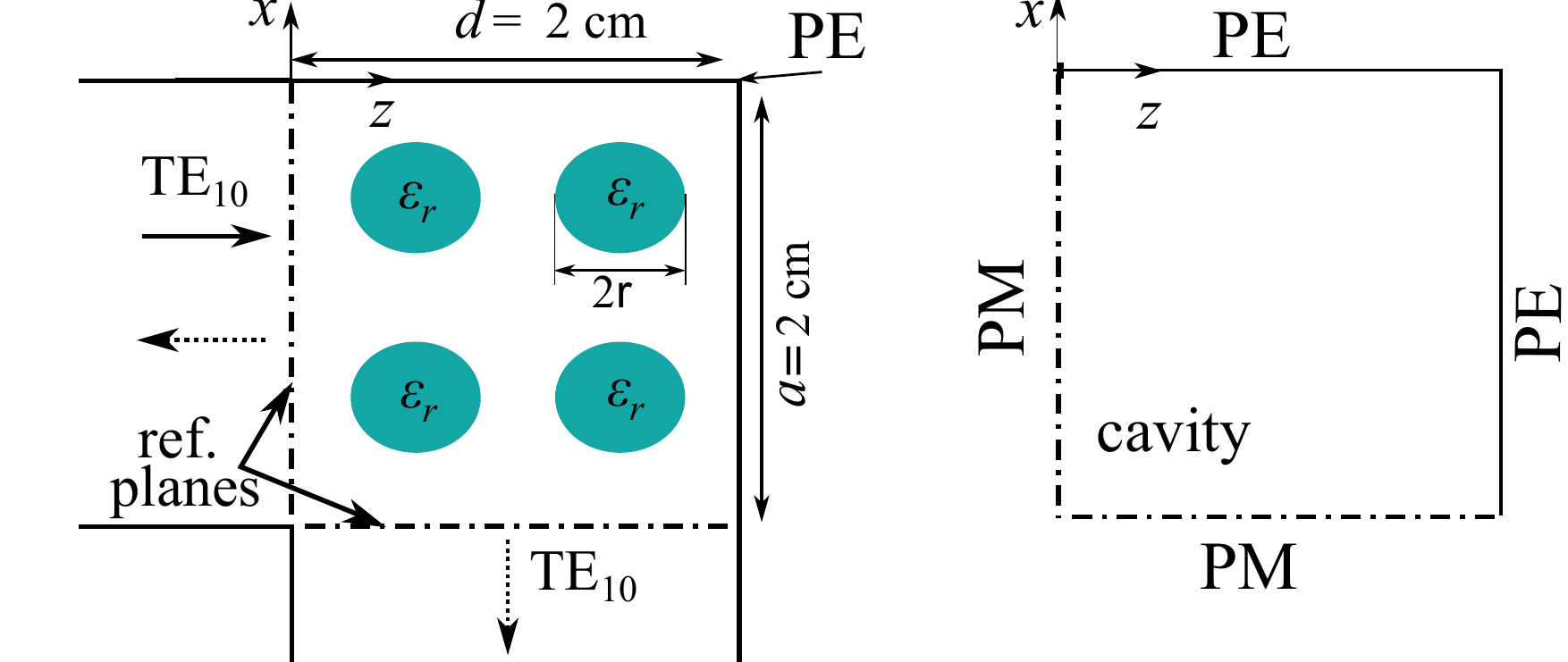}\caption{\label{nintybend}A $90^{0}$ waveguide bend loaded with a cylindrical
lossy dielectric post with constitutive parameters $\varepsilon_{r}=3,$
$\mu_{r}=1$, and $\tan\delta=0.01$ and the employed boundary conditions
for the canonical modes used by the EPT are shown on the right.}
\end{figure}

\begin{figure}
\centering \includegraphics[width=3in]{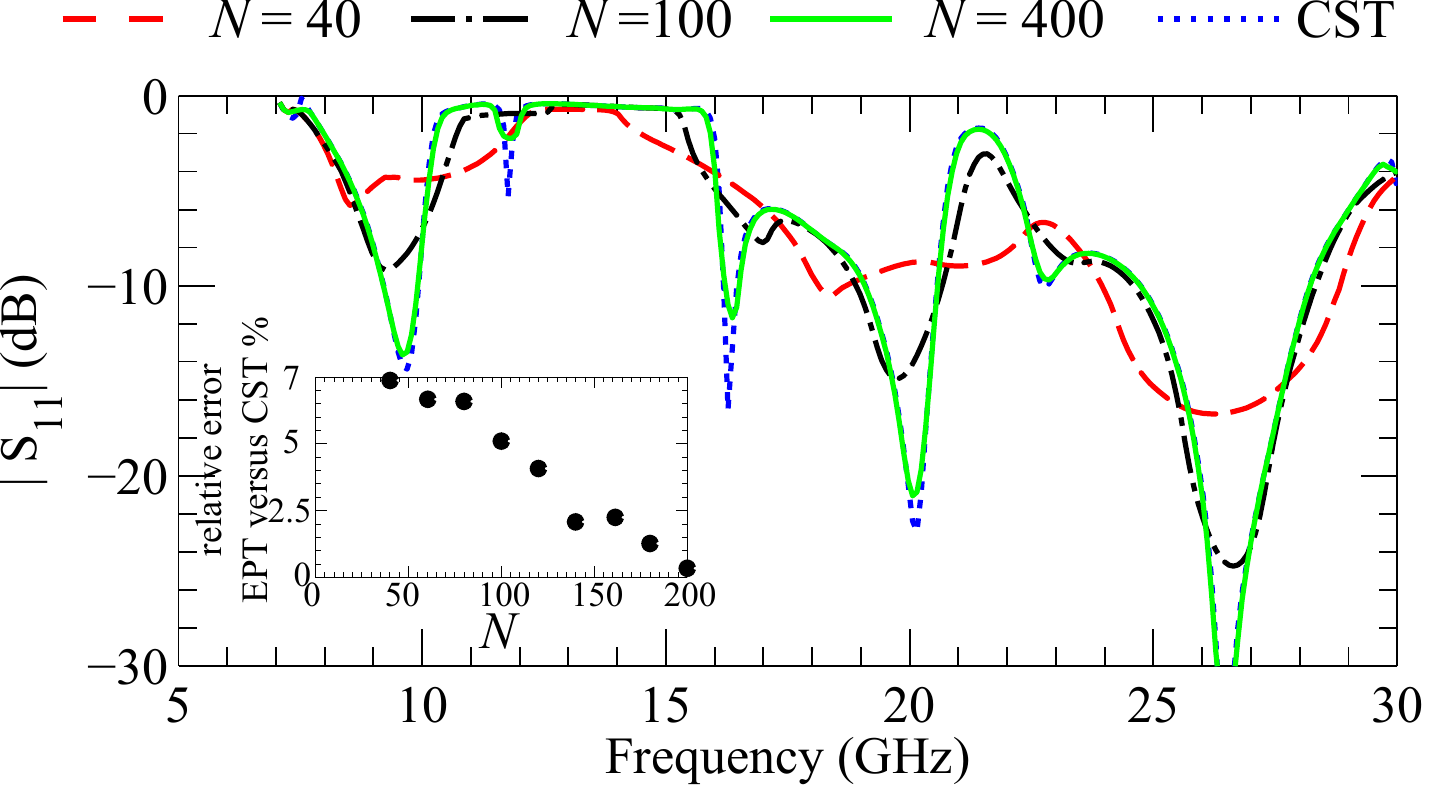} \label{f_90a} \\
 \centering \includegraphics[width=3in]{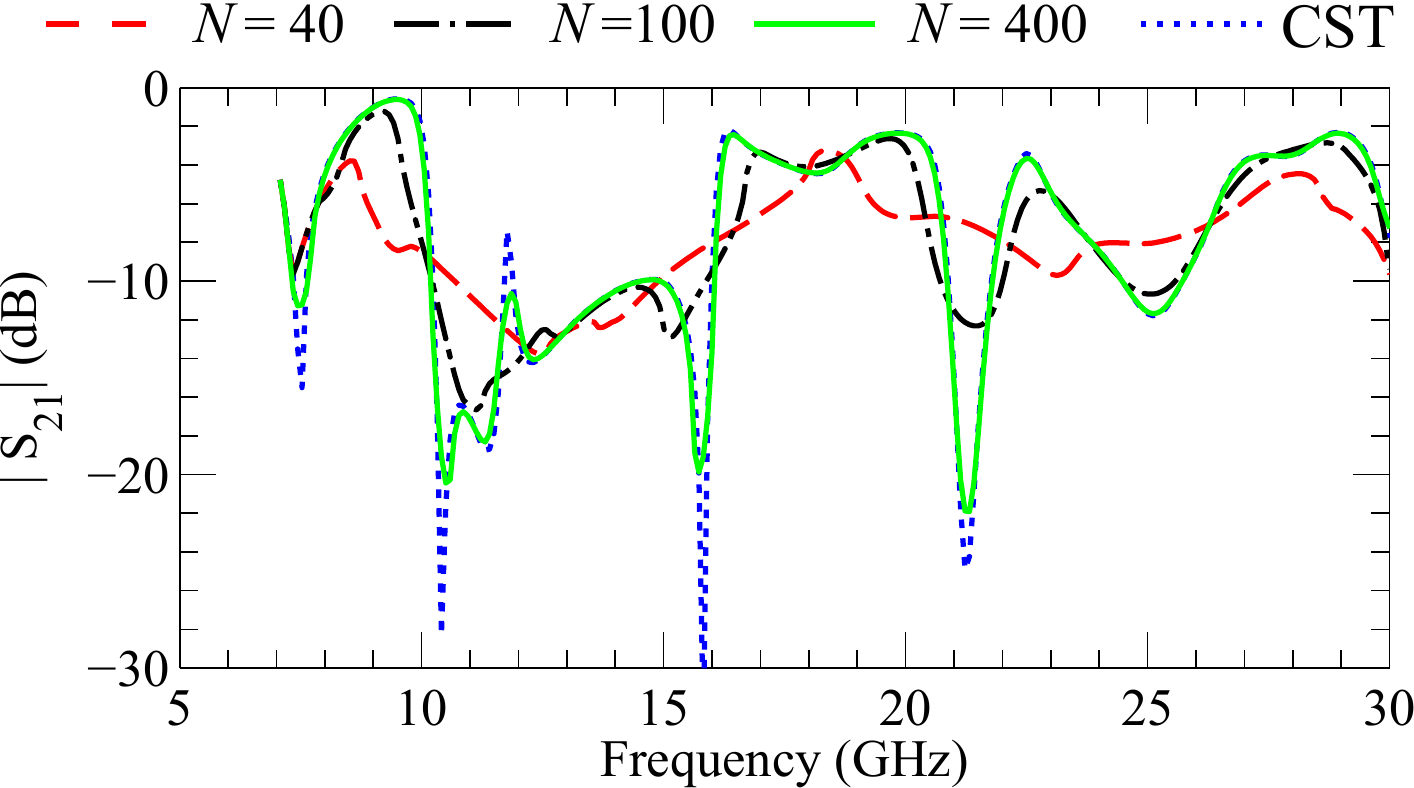} \label{f_90b} \caption{(a) $|S_{11}|$ and (b) $|S_{21}|$ of the loaded waveguide bend shown
in Fig. \ref{nintybend} computed using both EPT and full wave simulation
(CST). The inset of (a) shows the relative error between EPT calcualted
electric field in the cavity center compared to the one obtained using
CST, for increasing $N$.}
\label{ninety_results}
\end{figure}

The previous procedure is used here to solve an example of a $90^{\circ}$
rectangular waveguide bend shown in Fig. \ref{nintybend} having dimensions
$a=2$ cm and $d$$=2$ cm, partially loaded with four circular posts
with $\varepsilon_{r}=3$, $\mu_{r}=1$, $\tan\delta=0.01$ and of
radii $r=a/8$, and height of $b$ = 1 cm which is equal to that of
the waveguide in the y-direction. The input port is excited with the
$\textrm{TE}_{10}^{y}$ mode, and both ports are modeled as a PM surfaces.
The problem becomes y-invariant according to these particular choices,
i.e., a 2D problem (see Appendix for details of eigenmodes). Number
of mode harmonic variations along the x and z directions in the expansion
are taken equal, i.e. $N_{x}=N_{y}$. The scattering parameters
for the dominant mode are shown in Fig. \ref{ninety_results}, exhibiting
convergence for sufficient number of cavity eigenmodes and good agreement
with results obtained from full wave simulation (CST). 
It is worth nothing that all the EPT matrices can be computed analytically for shapes conforming with the chosen coordinate system,
therefore demonstrating robust performance for very complicated problems. In the inset of Fig. \ref{ninety_results}a, a plot
of the relative error between the electric field in the center of
the waveguide, i.e., at $x=a/2$, $z=d/2$ at 15 GHz calculated using
the EPT technique and the one obtained via CST is given, in which
good convergence for $N>200$ is observed. Note that the field inside
the dielectric would converge slower, however it is noticed that for
$N=400$ the maximum relative error between the field obtained using
EPT and the CST counterpart is less than 3\% indicating good overall
convergence.

\begin{figure}
\centering~\includegraphics[scale=0.5]{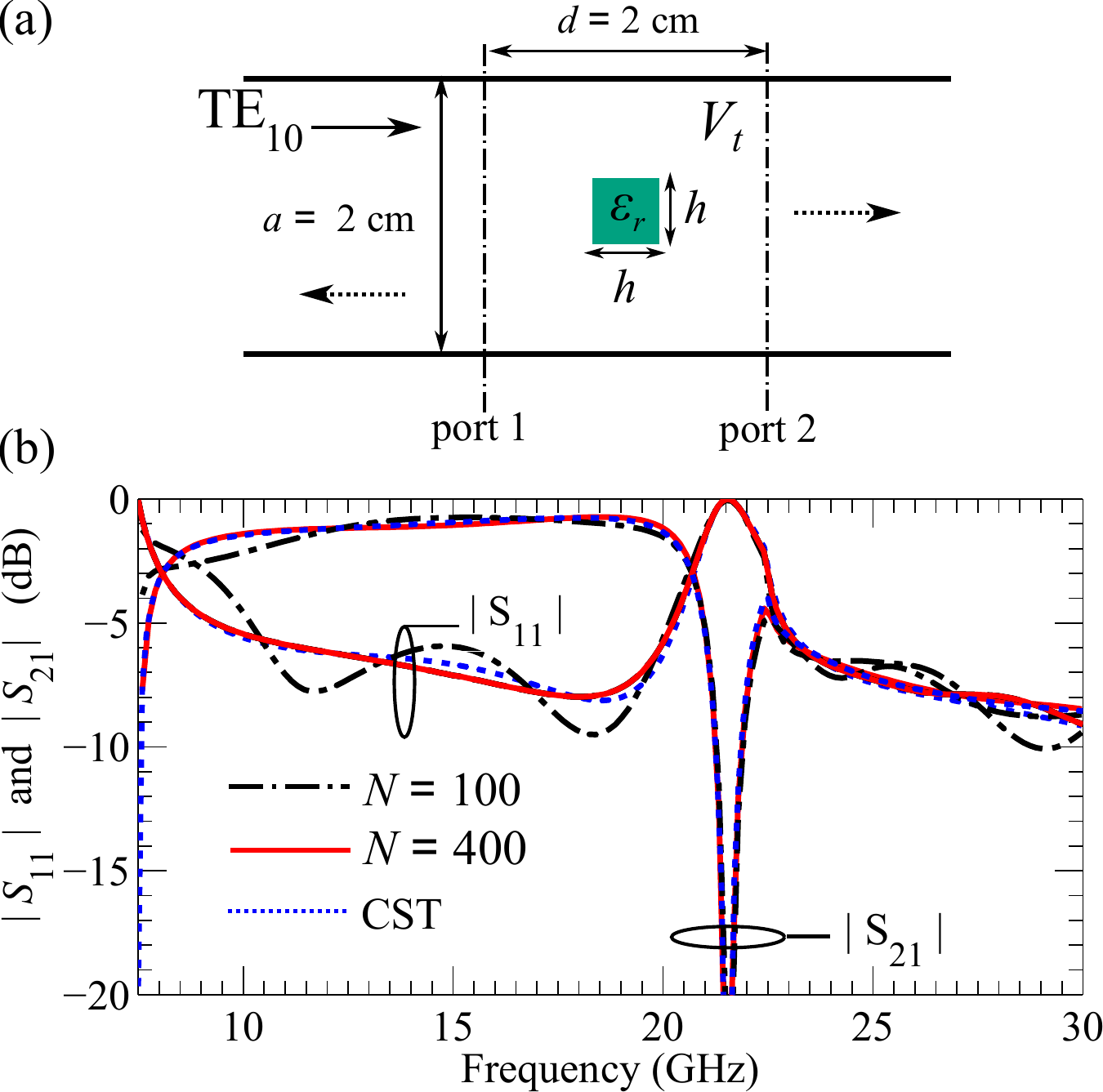} \centering

\caption{\label{S11-1-1}(a) A dielectric obstacle (square post) inside a rectangular
waveguide. (b) Scattering parameters of the loaded waveguide discontinuity
shown in (a) computed using both EPT and CST.}
\end{figure}

\begin{figure}
\centering
\begin{subfigure}{.2\textwidth}
\centering
\includegraphics[scale=0.25]{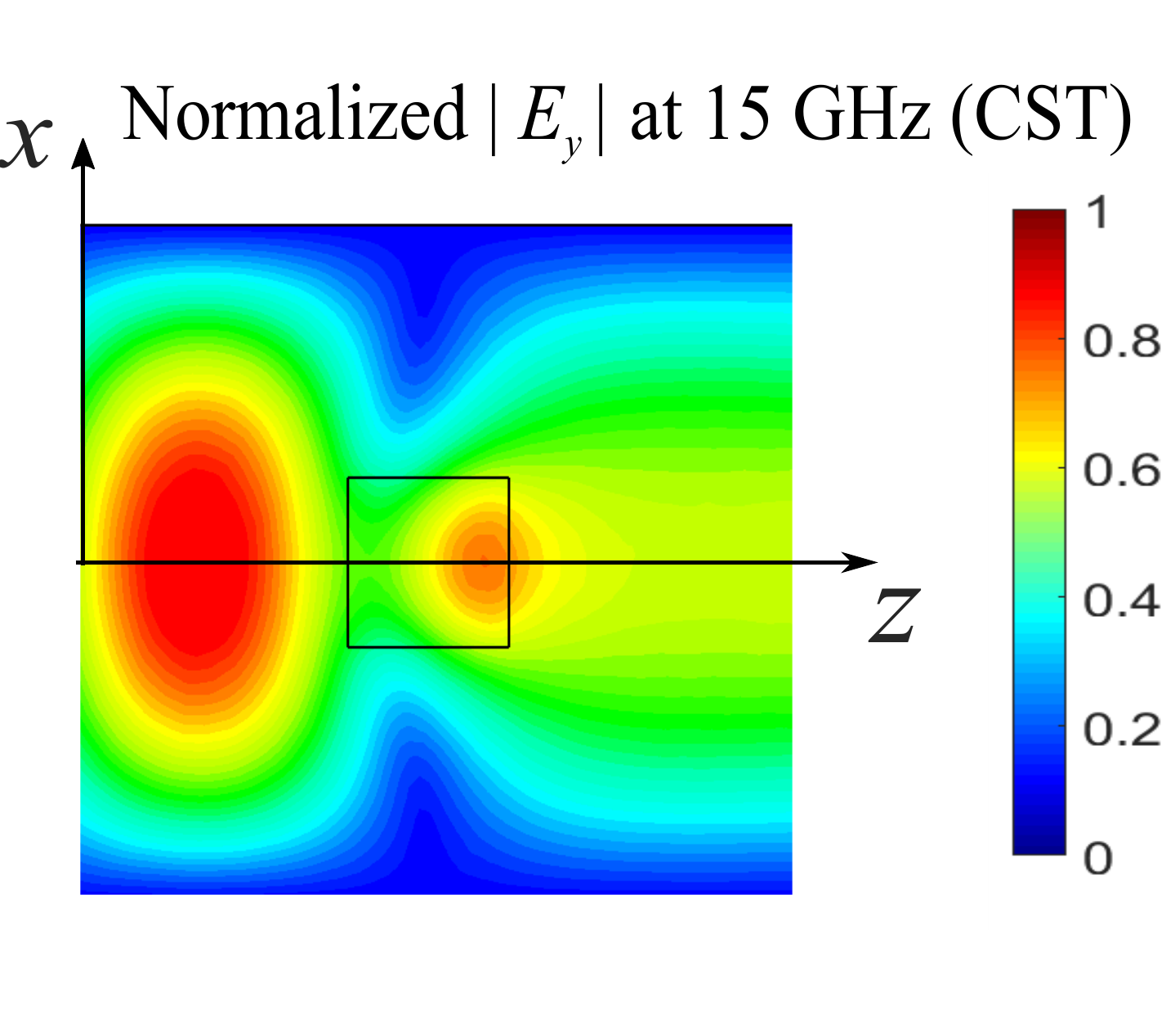}
\caption{}
\end{subfigure}
~
\begin{subfigure}{.2\textwidth}
\centering
\includegraphics[scale=0.25]{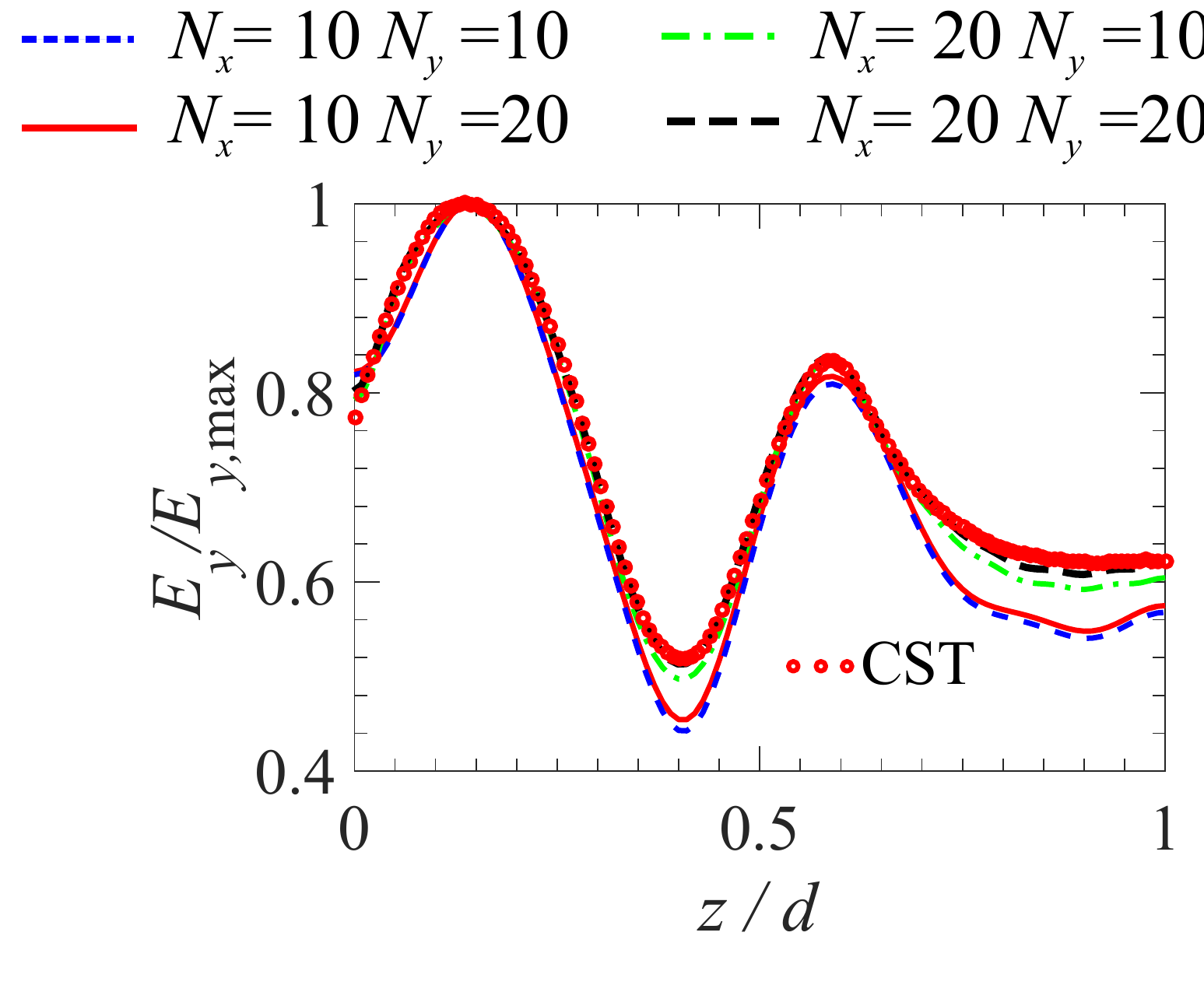}
\caption{}
\end{subfigure}
\\
\begin{subfigure}{.5\textwidth}
\centering
\includegraphics[scale=0.33]{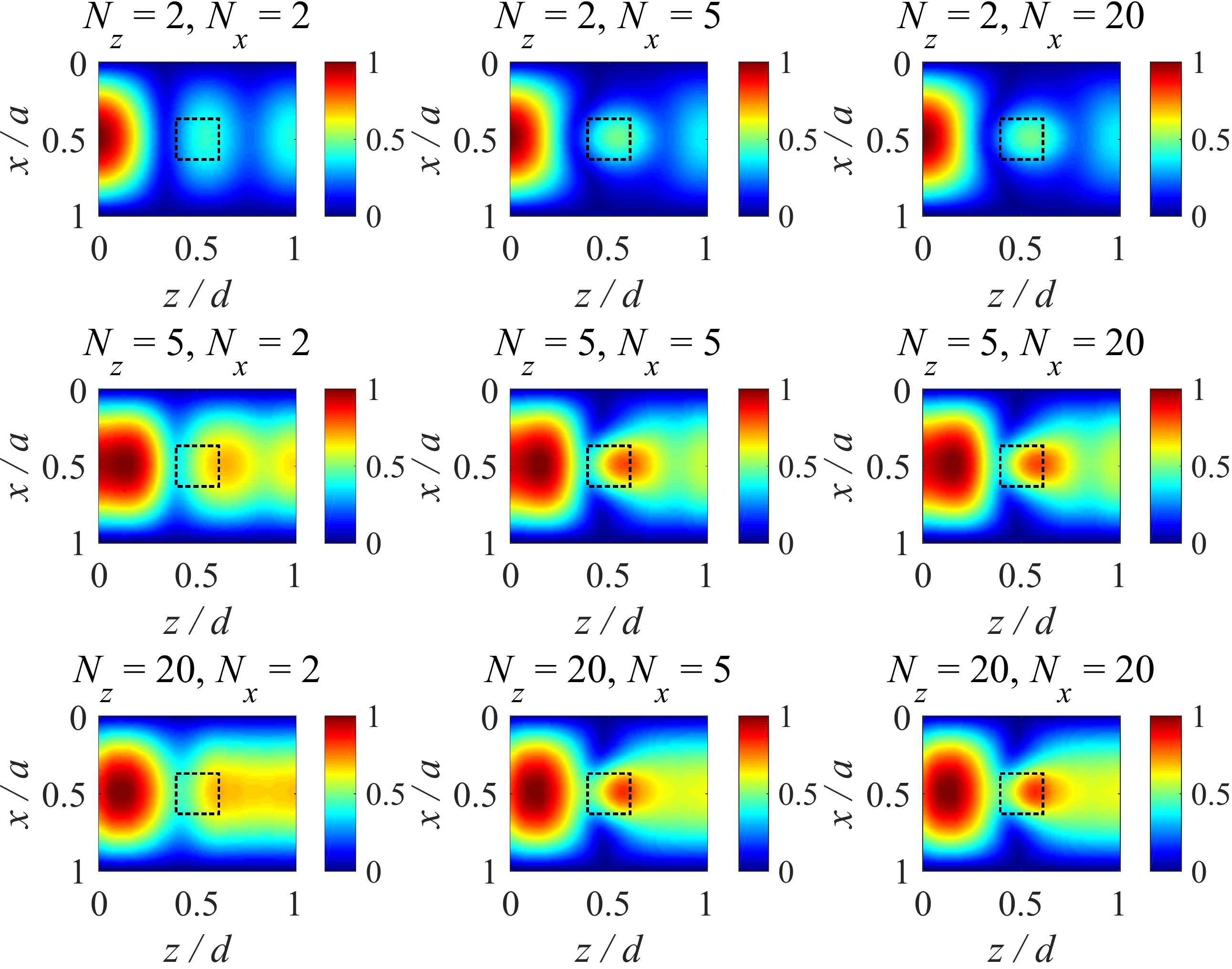}
\caption{}
\end{subfigure}
\caption{\label{Fieldocnvegence1}(a) Electric field magnitude |$E_{y}$|,
normalized ito its maximum inside the waveguide discontinuity in Fig.
\ref{S11-1-1} (a) at 15 GHz obtained by CST. (b) Field along the
z-direction for different N and showing a good convergence to that
obtained by CST. (c) Field maps of the y-directed electric field at
15 GHz obtained using the EPT technique. Here the field is excited
using the fundamental TE mode of the waveguide. }
\end{figure}

Another example of a dielectric post discontinuity in a rectangular
waveguide is shown in Fig. \ref{S11-1-1}(a), with a square post having
a dielectric constant $\varepsilon_{r}=3$ and side $h=0.5$ cm (the
post has also a height $b=1$ cm and located at the middle of the
lateral dimension $a$ of the waveguide). In Section III-C, the same geometry was used to demonstrate the EPT solution to resonance problem. Here, it is required to obtain the excited fields inside the
discontinuity due to the waveguide ports, and hence the scattering parameters. Good
agreement between the EPT and full-wave
simulation is achieved as seen in Fig. \ref{S11-1-1}(b) for $N$
= 400. Again, the convergence of the field distribution in the dielectric loaded waveguide section is investigated
at 15 GHz for dominant TE mode excitation. Fig.
\ref{Fieldocnvegence1}(a) provides the field distribution obtained using CST. The magnitude of the electric field $y$ component is plotted in Fig. \ref{Fieldocnvegence1}(b) along the $z$-axis showing convergence to the CST result for $N_{x}=N_{z}=20$,i.e., $N$
= 400. Also the convergence of the field in the $xz$ plane is illustrated in Fig. \ref{Fieldocnvegence1}(c) for increasing number of modes, approaching the plot of Fig. \ref{Fieldocnvegence1}(a) for $N=400$. It is worth noting here that the field converges steadily even for structures having sharp corners. Indeed, in all numerical methods, convergence of certain field components around sharp corners is relatively slow due to the edge behavior. This, however, does not necessarily affect the accuracy of the method or its well-posedness since the quantity of interest is typically related to the integral of the field/current distribution.

\begin{figure}
\centering{}\includegraphics[scale=0.5]{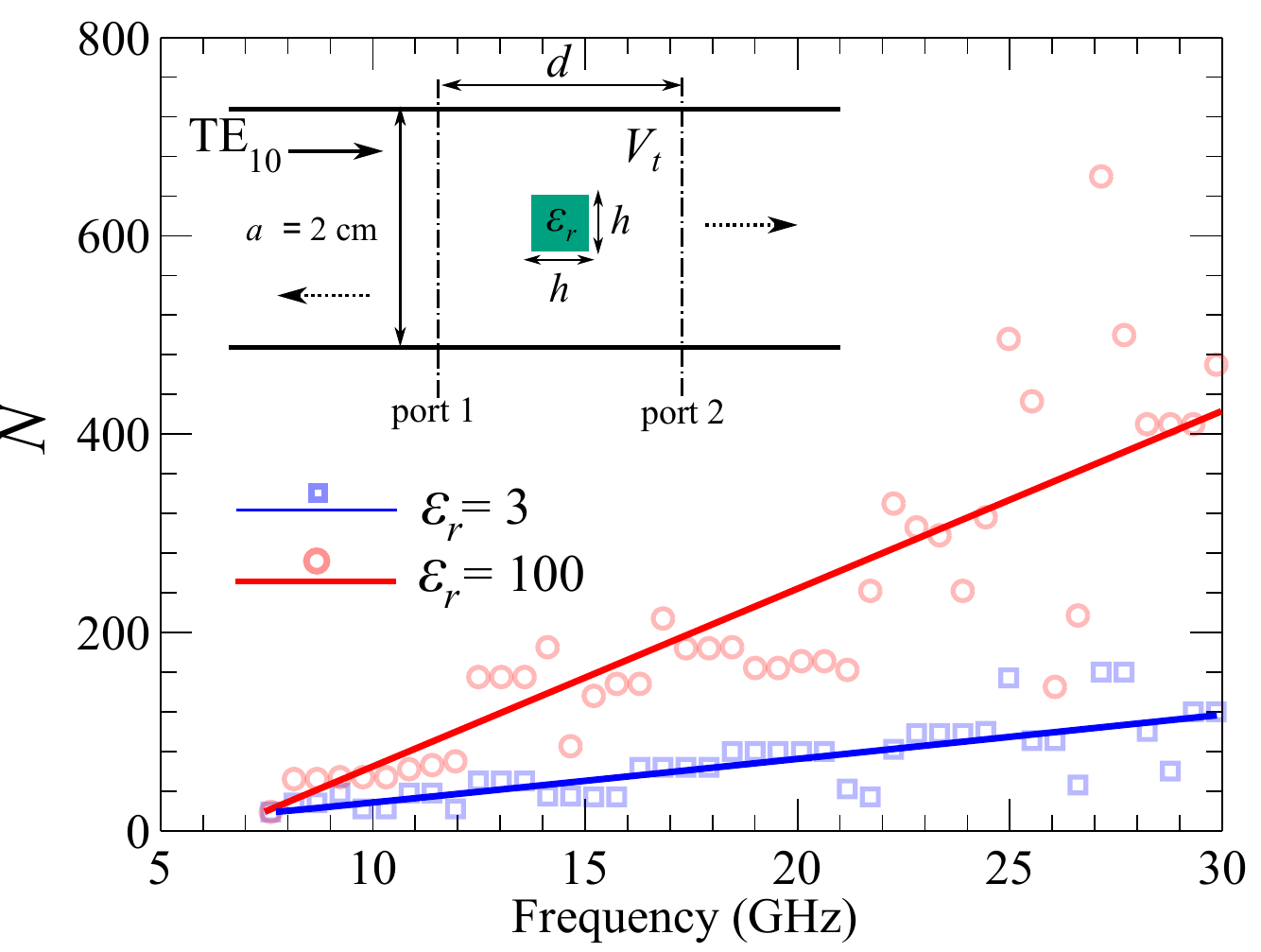}\caption{\label{epseffedctd}The number of expansion modes required to achieve
$\Delta S<1\%$ for $\varepsilon_{r}=$3 and 100 for the dielectric
discontinuity in the inset. The symbols represent the calculations
and the lines are the linear curve-fitting of $N$ with frequency
based on (\ref{ff}).}
\end{figure}

\begin{figure}
\centering{}\includegraphics[scale=0.39]{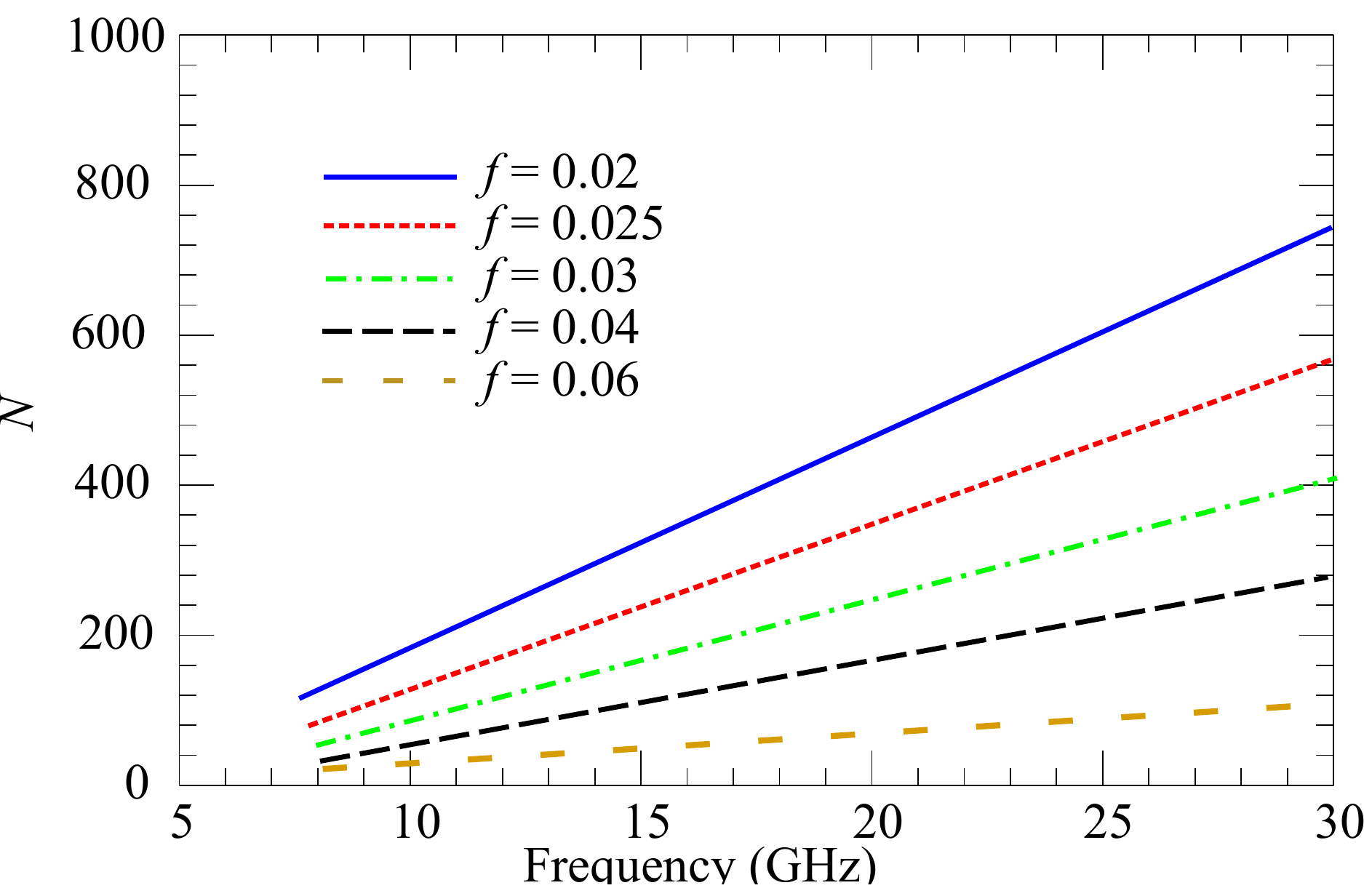}\caption{\label{conergvolume}Linear curve-fitting of the $N$ with frequency
for different values of the volumetric filling factor $f$ for the
case in the inset of Fig. \ref{epseffedctd}. }
\end{figure}

The purpose of the last example provided here is to determine a rule-of-thumb
for the number of eigenmodes necessary for convergence of the S-parameters
subject to variations in only one dimension of the structure. For
this aim, the number of modes required to reach a maximum relative
error in the magnitude of $|S_{11}|$ and $|S_{21}|$
less than 1\% in two consecutive iterations is computed, for two values for $\varepsilon_{r}$, and the results are shown in Fig. \ref{epseffedctd}. The required
number of modes (shown in Fig. \ref{epseffedctd} by faded symbols)
to achieve such criteria is increasing by increasing frequency as
well as increasing the dielectric constant of the filling material.
We also show a linear fitting of the number of modes such that $N\sim\omega$.
Not only the dielectric constant changes the requirement for convergence
but also the relative volume of the dielectric with respect to the
total cavity volume, or the dielectric filling factor of the dielectric
object defined as $f=V_{d}/V_{t}$ where $V_{d}$ is the volume of
the dielectric obstacle (in this example $V_{d}=h^{2}b$ and the filling
factor is reduced to $f=h/d$). This can be studied by varying the
value of the dimension $d$ of the cavity enclosure and keeping all
other dimensions constant, thus maintaining the same magnitude of
scattering parameters. For this case, the required value of $N$ to
reach less than a 1\% relative error in both $|S_{11}|$ and $|S_{21}|$)
for two consecutive passes is calculated, and a linear fitting of
such result is plotted against frequency as depicted in Fig. \ref{conergvolume}.
The results suggests that for smaller filling factor the convergence
is slower, i..e, $N$ increases as $f$ decreases. Based on these
linear fitting results, a simple rule-of-thumb may be developed for
the necessary number of modes, viz.
\begin{equation}
N>4k_{\textrm{eff}}D/f\label{ff}
\end{equation}
where $k_{\textrm{eff}}$ is the effective wavenumber $k_{\textrm{eff}}=\sqrt{\varepsilon_{r}f+(1-f)}k_{0}$,
$D$ is the largest dimension in the cavity enclosure, and here $f>0$.
Note that (\ref{ff}) is valid for variation in only one-dimension,
therefore for variations in three dimensions, analogous formula can
be postulated, which is left for future investigation.

It is worth mentioning that in
all simulations done using CST, mesh refinement is done such that
the error is less than 0.5\% in resonance frequency and the $S$ parameters
in two consecutive passes. Also, the number of port TE modes is $P=5$ in all simulations.

\section{Conclusion}

The eigenmode projection technique is demonstrated as a versatile
technique for solving electromagnetic cavity resonance and guided-wave
scattering problems. The solution procedure does not involve sub-wavelength
segmentation of the structure under consideration as with other numerical
techniques and can handle arbitrary-shaped structures with arbitrary
loading in a straightforward manner. Convergence studies were conducted
to estimate the number of modes required for the expansion to provide
accurate results for both resonance and scattering problems relative
to commercial solvers. Contrary to techniques like the Method of Moments,
the EPT does not require dealing with structures with known Green's
functions as it is based on defining a canonical cavity enclosing
the structure under study. Also, no singularity extraction process
is needed (as in Green's function based methods) nor divergence issues
are encountered (as in finite-differences based methods) during the
solution procedure. The EPT also gives the field distributions in
the quasi-analytical form of a weighted sum of simple eigenmodes (sinusoids
for rectangular cavities and Bessel functions for circular ones).
A few examples were presented to verify the proposed
technique in several scenarios. Other three-dimensional problems including
radiation problems, printed circuits and others are yet to be investigated.
For such problems, suitable choices of the canonical cavity, matrix
filling, and port modeling are to be developed.

\appendix

\begin{center}
CAVITY EIGENMODES\label{sec:Cavity-Eigenmode-Expansion}
\par\end{center}

The Helmholtz equations governing the solenoidal and irrotational
eigenmodes can be solved upon enforcing the boundary conditions on
the surface of the canonical cavity, which are
\begin{equation}
\text{For PM surface }\left\{ \begin{array}{c}
\mathbf{\hat{n}}\times\mathbf{H}_{n}=\mathbf{\hat{n}}\times\mathbf{G_{\lambda}}=\Psi_{\lambda}=\mathbf{0}\\
\mathbf{\hat{n}}\cdot\mathbf{E}_{n}=\mathbf{\hat{n}\cdot F_{\alpha}}=\partial\Phi_{\alpha}/\partial n=0
\end{array}\right.\label{eq:PM_BCs}
\end{equation}
\begin{equation}
\text{For PE surface }\left\{ \begin{array}{c}
\mathbf{\hat{n}}\times\mathbf{E}_{n}=\mathbf{\hat{n}}\times\mathbf{F_{\alpha}}=\Phi_{\alpha}=\mathbf{0}\\
\mathbf{\hat{n}}\cdot\mathbf{H}_{n}=\mathbf{\hat{n}\cdot G_{\lambda}}=\partial\Psi_{\lambda}/\partial n=0
\end{array}\right.\label{eq:PE_BCs}
\end{equation}

\subsection{Circular Cylindrical Cavity}

When solving problems illustrated in Figs. \ref{fig:cyl_cav_k}, \ref{fig:stepped_cav}
and \ref{fig:linac_cav}, where the structure under investigation
is a body of revolution around the $z$-axis, it is straight forward
that the most appropriate choice of the canonical cavity in this problem
is a finite-lenght circular cylinder with PEC walls. The cylinder
radius $a$ and length $d$ are chosen such that it completely enclose
the cavity. In this work, the solution is obtained for $\mathrm{TM^{z}}$
arbitrary cavity modes and the canonical cavity solenoidal and irrotational
eigenmodes will be derived.


The cavity solenoidal modes are easily obtained by expanding the two
curl equations in (\ref{eq:curl_solenoidal}) with $\partial/\partial\phi=0$.

The group of solenoidal magnetic field modes inside the solenoidal
cavity that will contribute to the ${\rm TM}{}_{z}$ mode excitation
can be obtained by choosing $\mathbf{H}_{n}=\nabla\times\Pi_{z}\hat{\mathbf{a}}_{z}$
where $\Pi$ is the Hertz potential, and solving (5) subject to the
boundary conditions (\ref{eq:PM_BCs})
and (\ref{eq:PE_BCs}), we get,
\begin{equation}
\mathbf{H}_{n}=H_{0n}J_{0}^{'}(k_{n_{1}}\rho)\cos(k_{n_{2}}z)\hat{\mathbf{a}}_{\phi},\label{eq:hphi_final-1}
\end{equation}
where $k_{n_{1}}=p_{0n_{1}}/a$, $k_{n_{2}}=n_{2}\pi/d$, and $k_{n}^{2}=k_{n_{1}}^{2}+k_{n_{2}}^{2}$;
and $p_{0}{n_{1}}$ is the $n_{1}$th zero of $J_{0}\left(x\right)$.
The constant $H_{0n}$ is determined such that the normalization condition
$\int_{V_{t}}\mathbf{H}_{n}\cdot\mathbf{H}_{n}dv=1$ is satisfied.

The corresponding $\mathbf{E}_{n}$ modes family can be obtained from
(\ref{eq:curl_solenoidal}),
\begin{multline}
\mathbf{E}_{n}=\frac{k_{n_{2}}}{k_{n}}H_{n}J_{0}^{'}\left(k_{n_{1}}\rho\right)\sin\left(k_{n_{2}}z\right)\hat{\mathbf{a}}_{\rho}\\
-\frac{k_{n_{1}}}{k_{n}}H_{n}J_{0}\left(k_{n_{1}}\rho\right)\cos\left(k_{n_{2}}z\right)\hat{\mathbf{a}}_{z}
\end{multline}

Similarly, the irrotational modes,$\Phi_{\alpha}$ and the corresponding
$\mathbf{F}_{\alpha}$, that couple to the azimothally symmetric $\textrm{TM}^{z}$
are obtained as,
\[
\Phi_{\alpha}=U_{\alpha}J_{0}\left(l_{\rho}\rho\right)\sin\left(l_{z}z\right)
\]
Hence, $\mathbf{F}_{\alpha}$ can be obtained from (\ref{eq:grad_irr}),
and $l_{\alpha}^{2}=l_{z}^{2}+l_{\rho}^{2}$. The constant $U_{\alpha}$
is determined such that the normalization condition $\int_{V_{t}}\Phi_{\alpha}^{2}dv=\int_{V_{t}}\mathbf{F}_{\alpha}\cdot\mathbf{F}_{\alpha}dv=1$
is satisfied.

\subsection{Rectangular Cavity}

When solving for the structures in Figs. \ref{nintybend} and \ref{epseffedctd},
the canonical cavity chosen would be a rectangular cavity resonator,
where the waveguide ports will be treated as PM surfaces when calculating
the cavity modes, while the rest of the walls is PE. As there is no
variation in the structure geometry in the normal direction ($y$-direction),
the excitation with $\textrm{TE}_{n0}^{z}$ would not couple to any
other port modes. Moreover, the excited cavity modes would be of the
type ${\rm TM_{n_{1}0n_{2}}^{y}}$.

The solenoidal electric field $\mathbf{E}_{n}$ modes for cavity in
Fig. \ref{nintybend} will be given by,
\[
\mathbf{E}_{n}=E_{n}\cos\left(k_{x}x\right)\cos\left(k_{z}z\right)\hat{\mathbf{a}}_{y},
\]
and the corresponding magnetic field $\mathbf{H}_{n}$ can be obtained
using Eq. (\ref{eq:curl_solenoidal}), where $k_{x}=\left(2n_{1}-1\right)\pi/\left(2a\right)$,
$k_{z}=\left(2n_{2}-1\right)\pi/\left(2d\right)$, and $k_{n}^{2}=k_{x}^{2}+k_{z}^{2}$.

The solenoidal electric field $\mathbf{E}_{n}$ modes for cavity in
Fig. \ref{epseffedctd} will be given by,
\[
\mathbf{E}_{n}=E_{n}\sin\left(k_{x}x\right)\cos\left(k_{z}z\right)\mathbf{a}_{y}
\]
and the corresponding magnetic field $\mathbf{H}_{n}$ can be obtained
using Eq. (\ref{eq:curl_solenoidal}), where $k_{x}=n_{1}\pi/2a$,
$k_{z}=n_{2}\pi/2d$, and $k_{n}^{2}=k_{x}^{2}+k_{z}^{2}$.

It is interesting to notice that irrotational modes will not couple
to any of the port excitation $\textrm{TE}_{n0}^{z}$ nor to any of
the previously mentioned ${\rm TM_{n_{1}0n_{2}}^{y}}$ solenoidal modes, for the special case of uniform posts along $y$-axis.
Moreover, few port modes are considered to account for evanescent
modes that are excited at the discontinuities.

\bibliographystyle{ieeetr}
\bibliography{MyReferences}

\end{document}